\documentclass[twocolumn,numberedappendix]{aastex62}

\usepackage{natbib}
\usepackage{multirow}
\usepackage{gensymb} 
\usepackage[flushleft]{threeparttable} 
\usepackage{amsmath}
\usepackage{url}
\usepackage{multirow}

\shorttitle{Stacked SEDs at $z\sim8$}
\shortauthors{Stefanon et al.}

\begin{document}

\title{Blue Rest-Frame UV-Optical Colors in $z\sim8$ Galaxies from GREATS: Very Young Stellar Populations at $\sim650$\, Myr of Cosmic Time}

\author{Mauro Stefanon}
\affiliation{Leiden Observatory, Leiden University, NL-2300 RA Leiden, Netherlands}

\author{Rychard J. Bouwens}
\affiliation{Leiden Observatory, Leiden University, NL-2300 RA Leiden, Netherlands}

\author{Ivo Labb\'e}
\affiliation{Centre for Astrophysics and SuperComputing, Swinburne, University of Technology, Hawthorn, Victoria, 3122, Australia}

\author{Garth D. Illingworth}
\affiliation{UCO/Lick Observatory, University of California, Santa Cruz, 1156 High St, Santa Cruz, CA 95064, USA}

\author{Pascal A. Oesch}
\affiliation{Departement d'Astronomie, Universit\'e de Gen\'eve, 51 Ch. des Maillettes, CH-1290 Versoix, Switzerland}
\affiliation{International Associate, Cosmic Dawn Center (DAWN), Niels Bohr Institute, University of Copenhagen and DTU-Space, Technical University of Denmark}

\author{Pieter van Dokkum}
\affiliation{Astronomy Department, Yale University, 52 Hillhouse Ave, New Haven, CT 06511, USA}

\author{Valentino Gonzalez}
\affiliation{Departamento de Astronom\'ia, Universidad de Chile, Casilla 36-D, Santiago 7591245, Chile}
\affiliation{Centro de Astrof\'isica y Tecnologias Afines (CATA), Camino del Observatorio 1515, Las Condes, Santiago 7591245, Chile}

\email{Email: stefanon@strw.leidenuniv.nl}

\begin{abstract}
Deep rest-optical observations are required to accurately constrain the stellar populations of $z\sim8$ galaxies.  Due to significant limitations in the availability of such data for statistically complete samples, observational results have been limited to modest numbers of bright or lensed sources.  To revolutionize the present characterization of $z\sim8$ galaxies, we exploit the ultradeep ($\sim27$ mag, $3\sigma$) \textit{Spitzer}/IRAC $3.6\mu$m and $4.5\mu$m data, probing the rest-frame optical at $z\sim8$, over $\sim200$ arcmin$^2$ of the GOODS fields from the recently completed GOODS Re-ionization Era wide-Area Treasury from Spitzer (GREATS) program, combined with observations in the CANDELS UDS and COSMOS fields. We stacked $\gtrsim100$ $z\sim8$ Lyman-Break galaxies in four bins of UV luminosity ($M_\mathrm{UV}\sim -20.7$ to $-18.4$) and study their $H_\mathrm{160}-[3.6]$ and $[3.6]-[4.5]$ colors. We find young ages ($\lesssim100$ Myr)  for the three faintest stacks, inferred from their blue $H_\mathrm{160}-[3.6]\sim 0$ mag colors, consistent with a negative Balmer break.  Meanwhile, the redder $H_\mathrm{160}-[3.6]$ color seen in the brightest stack is suggestive of slightly older ages. We explored the existence of a correlation between the UV luminosity and age, and find either no trend or fainter galaxies being younger. The stacked SEDs also exhibit very red $[3.6]-[4.5]\sim0.5$ mag colors, indicative of intense [\ion{O}{3}]+H$\beta$ nebular emission and SFR. The correspondingly high specific star-formation rates, sSFR$\gtrsim10$Gyr$^{-1}$, are consistent with recent determinations at similar redshifts and higher luminosities, and support the co-evolution between the sSFR and the specific halo mass accretion rate.
\end{abstract}


\section{Introduction}

The star-formation rate (SFR) and the stellar mass ($M_\star$) are two among the most fundamental physical parameters characterizing a galaxy. The SFR measures its recent (few millions to few hundred million years) rate of formation of new stars, while the stellar mass retains the cumulative effects of the (possibly varying) SFR over its entire life (i.e., its star-formation history - SFH), combined with its dark matter halo merger history. Comparing the evolution of the specific SFR (sSFR$\equiv$SFR$/M_\star$) to that of the specific halo mass accretion rate (SHMAR, $\equiv \dot{M_h}/M_h$, where $M_h$ is the dark matter halo mass, and the dot represents the time derivative) across cosmic time thus provides a useful diagnostic for the efficiency of the baryonic mass assembly, the hierarchical merging of the dark matter haloes, and the feedback mechanisms regulating the star-formation.

Programs such as GOODS (\citealt{giavalisco2004}), CANDELS (\citealt{grogin2011, koekemoer2011}), 3D-HST (\citealt{vandokkum2011, brammer2012, skelton2014, momcheva2016}), HFF (\citealt{lotz2017}), CLASH (\citealt{coe2013}), RELICS (\citealt{coe2019}) and the \textit{Hubble} Deep, Ultradeep and Extremely deep field (HDF/HUDF/XDF - \citealt{williams1996, beckwith2006,illingworth2013}) have allowed us to gain insights on, among other properties, the evolution of the sSFR up to $z\lesssim8$ (\citealt{gonzalez2010, labbe2013, stark2013, gonzalez2014, smit2014, salmon2015, faisst2016, davidzon2018}). A growing number of studies have found that the sSFR monotonically increases with increasing redshift (e..g, \citealt{duncan2014, faisst2016, davidzon2018}), with a factor $\sim10$ evolution  in the last $\sim10$\,Gyr of cosmic history (since $z\lesssim5-6$). 

The situation at higher redshifts is more uncertain. Observational evidence for a further factor $\sim5-10$ increase in the sSFR in the $\sim500$\,Myr between $5<z<8$ (e.g., \citealt{labbe2013, smit2014, salmon2015, faisst2016}) contrasts with a quasi-steady value reported by a number of authors (e.g., \citealt{marmol-queralto2016,santini2017, davidzon2018}). Furthermore, similarly discrepant results also exist for the evolution of the stellar-to-halo mass ratio.  Observational results have ranged from a constant $M_\star/M_h$ (see e.g., \citealt{durkalec2015, stefanon2017b, harikane2018}) to a $M_\star/M_h$ ratio evolving with redshift  (e.g., \citealt{finkelstein2015b, harikane2016}). Models show similar variation between an evolving $M_\star/M_h$ ratio (e.g., \citealt{behroozi2013, behroozi2019}) and modest evolution (e.g., \citealt{tacchella2018}).

These tensions could  arise at least in part  from systematic uncertainties in age estimates. While some recent work seems to suggest the existence of an age bimodality already at $z\gtrsim7$, with brighter galaxies characterized by more pronounced Balmer breaks (e.g., \citealt{jiang2016, castellano2017}),  there is also a recent indication of Balmer breaks in lower mass galaxies as well (e.g., \citealt{zheng2012, hoag2018, hashimoto2018, roberts-borsani2020}). These Balmer breaks perhaps result from more complex physical processes in place already in the first few hundreds Myr of cosmic history. Nevertheless, evidence indicative of young ages is seen from an increasing number of studies that point towards very strong emission lines at early cosmic times, with typical equivalent widths (EW) for the most abundant transitions (e.g., $H\alpha$, $H\beta$,  [\ion{O}{2}], [\ion{O}{3}], \ion{N}{2})  in excess of several hundred \AA ngstrom (e.g., \citealt{shim2011, labbe2013, stark2013, smit2014, smit2015, faisst2016, marmol-queralto2016, reddy2018, debarros2019, tran2020, endsley2021}). 

Given the lack of spectra for substantial samples of sources at $z\gtrsim5-6$, current age estimates rely on broad-band photometry bracketing the Balmer/4000\AA\ break (\citealt{bruzual1983, hamilton1985, balogh1999, kauffmann2003}), requiring measurements at these redshifts in the \textit{Spitzer}/IRAC bands. However, the relatively shallow coverage available from IRAC has not allowed us so far to probe the rest-frame optical of individual sources for significant samples of galaxies at $z\gtrsim 6-7$. 

Some new \textit{Spitzer}/IRAC observations allow us to revisit these issues for a \textit{HST} sample of high-redshift galaxies at $z\sim8$. In particular, we study the physical properties of a sample of star-forming galaxies at $z\sim8$ identified as Lyman-Break galaxies (LBGs) over four of the CANDELS fields (GOODS-N/S, UDS and COSMOS). The new dataset that enables this study is the recently completed  GOODS Re-ionization Era wide-Area Treasury from Spitzer (GREATS) program (PI: Labb\'e - \citealt{stefanon2021a}), which provides coverage in the $3.6\mu$m and $4.5\mu$m bands over $\sim200$\,arcmin$^2$ distributed over the GOODS-N and GOODS-S fields, to ultradeep limits of $\sim 27.1, 26.7$\,mag (AB, $5 \sigma$). This depth enables simultaneous detection in the $3.6\mu$m and $4.5\mu$m bands for an unprecedented $\sim40\%$ of the $z\sim8$ LBG sample in this study.

Our main results are based on stacked SEDs we construct from the parent sample of $z\sim8$ LBGs, that we slice into bins of UV luminosity and UV slope. We considered galaxies at redshift $z\sim8$ for two main reasons: 1) $z\sim8$ corresponds to the epoch of instantaneous re-ionization (\citealt{planck2016_reion}), thus probing the physical conditions of star formation in the heart of the re-ionization era and 2) At $7.2 \lesssim z \lesssim 9.0$ the [\ion{O}{3}]$\lambda\lambda4959,5007$ and $H\beta$ nebular lines lie within the $4.5\mu$m-band coverage, while  the $3.6\mu$m band is free from strong nebular emission ([\ion{O}{2}]$\lambda3727$ emission enters the $3.6\mu$m band with a smaller contribution to the broadband photometry - see Sect. \ref{sect:discussion}). This provides a solid reference for the measurements of the line intensities and, together with the flux in the $H$ band, for bracketing the ages of the stellar populations.

This paper is organized as follows. In Sect. \ref{sect:sample} we introduce the sample adopted in the present work which is based on \citet{bouwens2015}; in Sect. \ref{sect:stacking} we describe the procedure we followed to generate the stacked SEDs.  We present the stacked SEDs in Sect. \ref{sect:results}, and discuss our results in Sect. \ref{sect:discussion}. Finally, Sect. \ref{sect:conclusions} presents a summary and our conclusions. 

Throughout this paper, we adopt $\Omega_M=0.3$, $\Omega_\Lambda=0.7$ and $H_0=70$\,km s$^{-1}$ Mpc$^{-1}$, consistent with the most recent estimates from Planck (\citealt{planck2016_cosmology}). Magnitudes are given in the AB system (\citealt{oke1983}), while $M_\star$ and SFR refer to the \citet{salpeter1955} initial mass function (IMF). For brevity, we denote the \textit{HST} F435W, F606W, F775W, F850LP, F105W, F125W, F140W and F160W as $B_{435}$, $V_{606}$ and $i_{775}$, $z_{850}$, $Y_{105}$, $J_{125}$, $JH_{140}$ and $H_{160}$.

\section{Data set and sample selection}
\label{sect:sample}

\begin{figure}
\hspace{-0.5cm}\includegraphics[width=9cm]{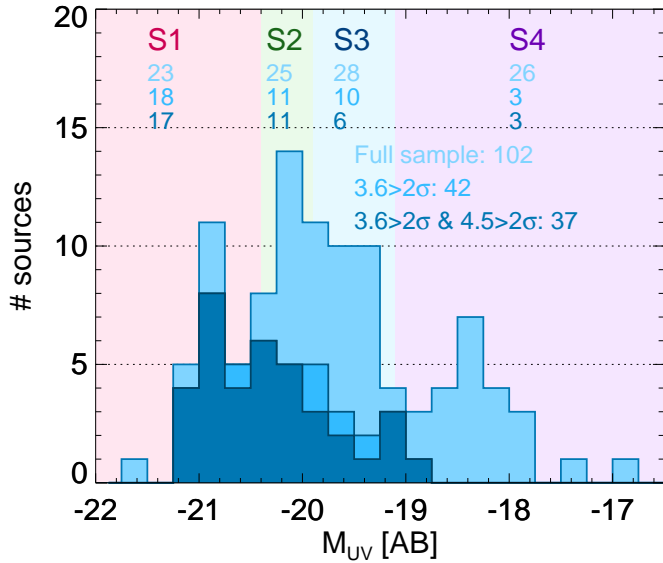}
\caption{Distribution of the rest-frame UV absolute magnitudes of the sample of $z\sim8$ candidate LBGs adopted for this work.  The four colored background regions correspond to the four bins in absolute magnitude we adopted in our stacking, as labelled at the top. S1 includes  galaxies  with $M_\mathrm{UV} < -20.4$\, mag,  S2 galaxies with  $-20.4 < M_\mathrm{UV} <-19.9$, S3 galaxies with $-19.9 < M_\mathrm{UV} <-19.1$, while S4 includes  galaxies  $M_\mathrm{UV} > -19.1$\,mag. This binning scheme was chosen to evenly distribute the number of sources across the four bins and enable the study of trends with luminosity. The different shades of blue correspond to different cuts in IRAC detection significance, as indicated by the legend, where we also quote each sample size. The break-down of these sizes in each $M_\mathrm{UV}$ bin is presented by the three numbers in column below the stack labels.  \label{fig:sample}}
\end{figure}

\begin{deluxetable*}{llccccc}
\tablecaption{Observational data used for the SMF estimates. \label{tab:obs_data}}
\tablehead{\multicolumn{2}{c}{Field} & \colhead{Area} & \colhead{$H_\mathrm{160}$\tablenotemark{a}} &\colhead{IRAC Data\tablenotemark{b}} & \colhead{$3.6\mu$m\tablenotemark{c}} & \colhead{$4.5\mu$m\tablenotemark{c}}    \\
\multicolumn{2}{c}{Name} & \colhead{[arcmin$^2$]} & \colhead{$5\sigma$ [mag]} & & \colhead{$5\sigma$ [mag]} & \colhead{$5\sigma$ [mag]}  
}
\startdata
\multicolumn{2}{l}{XDF}      & $4.7$   & $29.4$ & GREATS        & $\sim27.2$  & $\sim26.7$   \\
\multicolumn{2}{l}{HUDF09-1} & $4.7$   & $28.3$ & GREATS        & $\sim26.3$  & $\sim25.8$   \\
\multicolumn{2}{l}{HUDF09-2} & $4.7$   & $28.7$ & GREATS        & $\sim27.0$  & $25.5-26.0$  \\
\multicolumn{2}{l}{ERS}      & $40.5$  & $27.4$ & GREATS        & $26.2-27.0$ & $25.6-26.7$  \\
CANDELS & GOODS-N Deep       & $62.9$  & $27.5$ & GREATS        & $27.0-27.3$ & $26.5-26.8$  \\
        & GOODS-N Wide       & $60.9$  & $26.7$ & GREATS        & $26.3-27.2$ & $25.8-26.8$  \\
        & GOODS-S Deep       & $64.5$  & $27.5$ & GREATS        & $\sim27.3$  & $26.6-26.9$  \\
        & GOODS-S Wide       & $34.2$  & $26.8$ & GREATS        & $26.5-27.2$ & $26.2-26.7$  \\
        & COSMOS             & $151.9$ & $26.8$ & SEDS+SCANDELS & $26.4-26.7$ & $26.0-26.3$  \\
        & UDS                & $151.2$ & $26.8$ & SEDS+SCANDELS & $25.4-26.3$ & $25.0-25.9$  \\[5pt]
\hline
\multicolumn{2}{c}{Totals:} & $580.2$ & & & &  \\
\enddata
\tablenotetext{a}{$5\sigma$ limit from \citet{bouwens2015}, computed from the median of measured uncertainties of sources.}
\tablenotetext{b}{GREATS: \citet{stefanon2021a}; SEDS: \citet{ashby2013a}; SCANDELS: \citet{ashby2015}.}
\tablenotetext{c}{Nominal $5\sigma$ limit for point sources from the SENS-PET exposure time calculator, based on the exposure time maps. Due to inhomogeneities in the coverage, a range of values is quoted when the depth varies by more than $\sim0.2$\,mag across the field. Because of the combined effects of broad \textit{Spitzer}/IRAC PSF and significant exposure times, source blending may reduce the actual depth (see discussion in \citealt{labbe2015}).}
\end{deluxetable*}

We adopted the sample of $Y$ dropouts assembled by \citet{bouwens2015} identified over the CANDELS (\citealt{grogin2011, koekemoer2011}) GOODS-N, GOODS-S (\citealt{giavalisco2004}), EGS (\citealt{davis2007}), UDS (\citealt{lawrence2007}) and COSMOS (\citealt{scoville2007}) fields, the ERS field (\citealt{windhorst2011}), and the UDF/XDF (\citealt{beckwith2006,illingworth2013, ellis2013}) with the HUDF091- and HUFD09-2 parallels (\citealt{bouwens2011b})\footnote{We excluded CANDELS/EGS because of the lack of deep data in the Y band, which makes the selection of candidate $z\sim8$ LBGs more uncertain.}.  The data over the CANDELS UDS and COSMOS fields, although shallower, provide wide area coverage critical for sampling $L\gtrsim L^*$ galaxies. Table \ref{tab:obs_data} summarizes the main properties of the adopted data sets.

The COSMOS and UDS fields lack coverage in the \textit{HST}/WFC3 $Y_{105}$ band, key for a robust selection of $z\sim8$ sources. We therefore complemented the \textit{HST} photometry in these fields with measurements from  the deep ground-based NIR mosaics of the UltraVISTA DR3 \citep{mccracken2012} and VIDEO (\citealt{jarvis2013}) programs, respectively. Additionally, these two fields benefit from  deep public Subaru/SuprimeCam imaging in the $z$ band. Overall, these data have $5 \sigma$ depths in the $Y$ band ranging from $26.7-27.5$~mag (GOODS fields) to $\sim26.0$~mag (UDS and COSMOS), $\sim26.8-27.8$ in the $J$ and $H$ bands and about $27.5$~mag in the optical bands.

Photometry on the ground-based data was performed with \textsc{Mophongo} (\citealt{labbe2006, labbe2010a, labbe2010b, labbe2013, labbe2015}), which uses the brightness profile of each source from a high-resolution image (a combination of $J_{125}$, $JH_{140}$ and $H_{160}$ in our case) to remove the corresponding neighbouring objects before performing aperture photometry. For the aperture photometry we adopted an aperture of $1\farcs2$ diameter, and corrected to total magnitudes using the brightness profile of each source on the low-resolution image and the point-spread function (PSF) reconstructed at the specific locations of each source.

The candidate $z\sim8$ Lyman-Break galaxies (LBGs) were initially selected as $Y$-band dropouts, with the additional constraint of non-detection at $2 \sigma$  in each of  the bands bluer than $Y$. Specifically, the following color criteria were adopted (see  \citealt{bouwens2015} for details):
\begin{equation}
\begin{split}
(Y_{105} - J_{125} > 0.45) &\land (J_{125} - H_{160} < 0.5) \\
&\land \\
(Y_{105} - J_{125}) > 0.75&(J_{125} - H_{160}) +0.525 
\end{split}
\end{equation}
where $\land$ denotes the logical AND operator. These criteria resulted in an initial sample of $185$ sources. 

To gain information on the rest-frame optical properties of the bona-fide $z\sim8$ sources, we incorporated photometry in the \textit{Spitzer/}IRAC \citep{fazio2004} $3.6\mu$ and $4.5\mu$m bands from the SEDS (\citealt{ashby2013}) and  S-CANDELS (\citealt{ashby2015}) programs for the COSMOS and UDS fields,  and from the \textit{GOODS Re-ionization Era wide-Area Treasury from Spitzer} (GREATS; \citealt{stefanon2021a}) for sources in the GOODS-N/S fields. GREATS (PI: I. Labb\'e) is a recently completed \textit{Spitzer} legacy program which brings near-homogeneous $\sim200-250$~hr depth (corresponding to $5 \sigma \sim 26.8-27.1$~mag) in \textit{Spitzer}/IRAC $3.6\mu$m and $4.5\mu$m for 200 arcmin$^2$ over the two GOODS fields.  Flux densities in the IRAC bands were derived using the same \textsc{Mophongo} tool, adopting an aperture of $1\farcs8$ diameter.

\begin{figure}
\hspace{-0.5cm}\includegraphics[width=9cm]{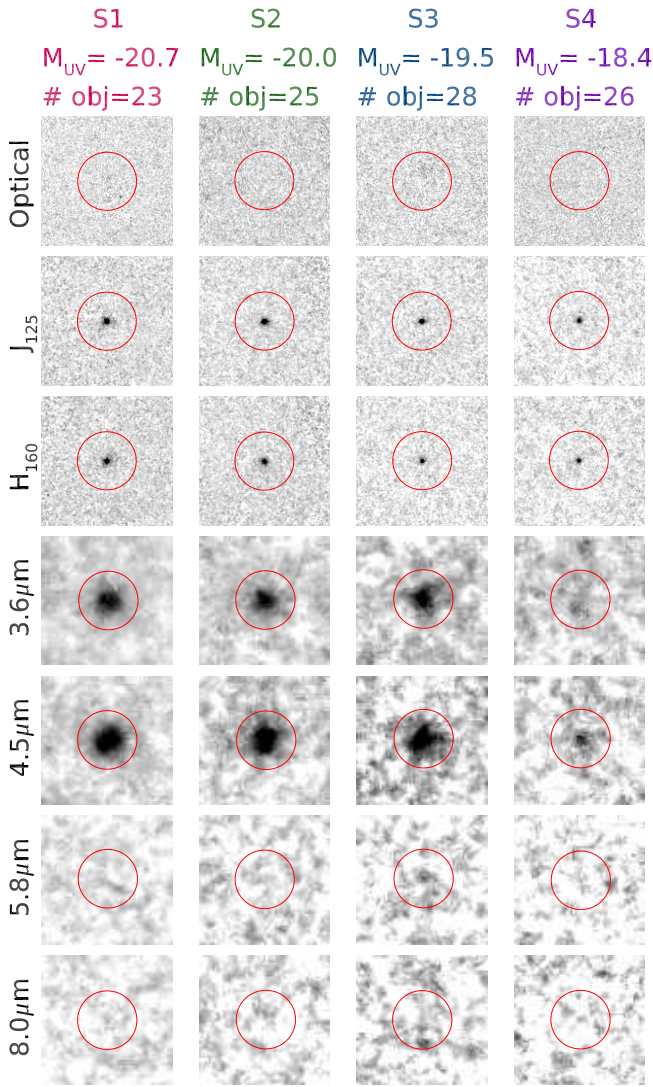}
\caption{Image stamps of the IRAC and \textit{HST} median stacks. Each cutout is $\sim8\farcs0\times8\farcs$ in size. The red circle marks the $3\farcs6$ diameter aperture adopted for the photometry of the IRAC stacks. \textit{HST} stacks are presented to provide a better visual context of the data involved in our study, as the median flux densities in \textit{HST} bands were estimated from the photometry of individual sources. Each row refers to a different band, as labeled on the left side; in particular the \textit{HST} optical stack combines all data available in the $B_{435}, V_{606}, i_{775}$ and $z_{850}$ bands. The columns refer to a stack in a specific luminosity bin, as labeled at the top. The increased coverage offered by the GREATS program enables detections in the $3.6\mu$m and $4.5\mu$m bands to faint limits ($M_\mathrm{UV}=-19.5$). Remarkably, we find a $\gtrsim2\sigma$ detection for the flux density in the $4.5\mu$m band for the faintest luminosity bin (S4, $M_\mathrm{UV}=-18.4$) and a $\sim1.4\sigma$ measurement in the $3.6\mu$m band. The $5.8\mu$m and $8.0\mu$m-band stacks, instead, are characterized by non detections ($< 2 \sigma$)  irrespective of their UV luminosity. \label{fig:irac_stacks}}
\end{figure}

We computed photometric redshifts and rest-frame UV$_{1600}$ luminosities with \textsc{EAzY} \citep{brammer2008}, complementing its default template set with three  2.5~Gyr passively evolving, $Z_\odot$, \citealt{bruzual2003} templates with extinction $A_\mathrm{V}=2.0, 5.0$ and $8.0$ mag (\citealt{calzetti2000}). Recent studies of high-redshift galaxies have revealed that strong nebular emission are ubiquitous  (e.g., $H\alpha$, $H\beta$,  [\ion{O}{2}], [\ion{O}{3}], \ion{N}{2} - e.g., \citealt{shim2011, labbe2013, stark2013, smit2014, smit2015, faisst2016, marmol-queralto2016, reddy2018, debarros2019, tran2020, endsley2021}). At $z\sim8$ the \textit{Spitzer/}IRAC $3.6\mu$m band is potentially contaminated by [\ion{O}{2}], the $4.5\mu$m band  by $H\beta$ and [\ion{O}{3}],  and the $5.8\mu$m band by $H\alpha$ and \ion{N}{2}. Therefore, to minimize potential systematic biases in our $z\sim8$ sample selection, and in the average properties of the sample, we excluded the IRAC bands when computing the photometric redshifts. For sources to be included in our sample we then required  that their photometric redshift be $7.3 \le z_\mathrm{phot} \le 8.7$. We selected this range to allow for modest redshift errors in localizing the [\ion{O}{3}]$_{4959,5007}$ lines to the $4.5\mu$m band.  Finally, we visually inspected the neighbour-cleaned cutouts of each source in the IRAC $3.6\mu$m and $4.5\mu$m bands and excluded from the sample those with residuals from nearby \textsc{Mophongo}-subtracted IRAC sources overlapping with the nominal position of the source. We also excluded those where the background showed gradients, since both issues could introduce systematic biases during the stacking process. 

After applying the above criteria, the final sample of candidate $z\sim8$ LBGs included $102$ sources. Their distribution in absolute magnitude is presented in Figure \ref{fig:sample}. The heterogeneity of our carefully derived data sets allows us to probe a broad interval in UV luminosities, ranging from $\gtrsim L^*$ to $\sim0.06L^*$. Remarkably, $\sim40\%$ of the sources are detected in the $3.6\mu$m band with a significance in excess of $2\sigma$.

\section{Stacking procedure}
\label{sect:stacking}
Our main results are based on stacking analysis,  allowing us to probe the average physical properties of LBGs at $z\sim8$, while self-consistently including in our analysis all those sources with limited  detection significance in one or more bands.

In our analysis we segregate sources in bins of luminosity (irrespective of UV slope) and of UV slope (irrespective of luminosity). However, in both cases we applied the same stacking procedure described below.

\begin{deluxetable*}{lr@{ $\pm$}rr@{ $\pm$}rr@{ $\pm$}rr@{ $\pm$}r}
\tablecaption{Flux densities for the four stacked luminosity-selected SEDs \label{tab:photometry}}
\tablehead{\colhead{Filter} & \twocolhead{S1} & \twocolhead{S2} & \twocolhead{S3}& \twocolhead{S4}  \\
& \twocolhead{[nJy]} & \twocolhead{[nJy]} & \twocolhead{[nJy]} & \twocolhead{[nJy]} 
}
\startdata
    $B_\mathrm{435}$ & \multicolumn{2}{c}{$$}  & $     0.4$ & $    2.6$  &  \multicolumn{2}{c}{$$}  &  \multicolumn{2}{c}{$$}  \\
    $V_\mathrm{606}$ & $    -0.3$ & $    3.1$  & $     0.0$ & $    1.9$  & $    -2.1$ & $    1.6$  & $    -0.3$ & $    0.5$  \\
    $i_\mathrm{775}$ &  \multicolumn{2}{c}{$$}  & $    -2.3$ & $    3.1$  & $    -0.8$ & $    2.3$  & $    -0.2$ & $    0.7$  \\
    $z_\mathrm{850}$ &  \multicolumn{2}{c}{$$}  & $    -1.0$ & $    3.7$  & $    -1.7$ & $    2.3$  & $     0.1$ & $    1.1$  \\
    $Y_\mathrm{105}$ &  \multicolumn{2}{c}{$$}  &  \multicolumn{2}{c}{$$}  & $    14.3$ & $    3.3$  & $     4.6$ & $    0.9$  \\
    $J_\mathrm{125}$ & $   101.4$ & $    6.5$  & $    55.7$ & $    3.1$  & $    35.1$ & $    1.9$  & $    12.5$ & $    0.8$  \\
    $H_\mathrm{160}$ & $    99.8$ & $    6.3$  & $    50.0$ & $    3.8$  & $    32.1$ & $    2.3$  & $    10.6$ & $    0.8$  \\
      IRAC $3.6\mu$m & $   118.2$ & $   24.2$  & $    47.3$ & $    9.7$  & $    25.5$ & $    6.4$  & $     6.7$ & $    4.8$  \\
      IRAC $4.5\mu$m & $   174.7$ & $   24.9$  & $    85.1$ & $   12.8$  & $    43.6$ & $    8.6$  & $    16.0$ & $    7.3$  \\
      IRAC $5.8\mu$m & $   207.1$ & $  167.3$  & $    43.2$ & $   53.4$  & $   126.2$ & $   67.1$  & $     3.1$ & $   61.2$  \\
      IRAC $8.0\mu$m & $    -8.6$ & $  223.9$  & $   -14.8$ & $   83.0$  & $   139.3$ & $   86.6$  & $   -48.7$ & $   68.0$  \\
\enddata
\tablecomments{Measurements for the ground-based and \textit{Spitzer/}IRAC bands are $1\farcs2$ aperture fluxes from \textsc{mophongo} corrected to total using the PSF and luminosity profile information; \textit{HST/}WFC3-band flux densities are measured in $0\farcs6$ apertures and converted to total using the PSF growth curves. Measurements involving $<90\%$ of sources have been left blank. We omit $JH_\mathrm{140}$ because its coverage is available for $<90\%$ of sources for any stack.}
\end{deluxetable*}

For the analysis in bins of luminosity, we chose the thresholds in UV magnitudes as a tradeoff between maximising the number of sources in each bin, improving the S/N in the IRAC $3.6\mu$m and $4.5\mu$m bands, and enabling to inspect potential correlations of the results with UV luminosity. Hereafter we will refer to each bin as S1, S2, S3 and S4 in order of decreasing UV luminosity. Specifically, S1 includes galaxies more luminous than $M_\mathrm{UV}=-20.4$ mag, S2 galaxies with $-20.4<M_\mathrm{UV}<-19.9$, S3 galaxies with $-19.9<M_\mathrm{UV}<-19.1$, while S4 includes galaxies fainter than $M_\mathrm{UV}=-19.1$. These luminosity bins were chosen as a trade-off between maximizing the number of sources in each bin and having a sufficient number of bins to quantify possible dependencies on luminosity. The UV magnitude distribution of sources in the four bins is presented in Figure \ref{fig:sample}.

For the analysis in bins of UV slope, we divided the sample in five bins, centered at UV slopes $\beta=-2.5, -2.3, -2.1, -1.9$ and $-1.7$, with a width of $\Delta\beta=0.2$. Since the stacking procedures were similar for both samples, and the luminosity-binned stacks are the primary focus of this paper, the discussion that follows is for the luminosity-selected stacks.

To further mitigate the dependence on the brightness of individual sources within each bin, before deriving the actual stacked fluxes, we normalized the flux density in each band and for each source with the corresponding inverse-variance weighted average of the $J_{125}$ and $H_{160}$ flux densities. The final values were obtained by taking the median of the measurements. Uncertainties were computed bootstrapping the sample 1000 times, drawing with replacement the same number of objects in the considered bin and taking the standard deviation of the measurements to be the flux uncertainty after perturbing the fluxes according to their photometric uncertainty (see also e.g., \citealt{gonzalez2011, stefanon2017b}). This procedure also allows us to naturally account for the intrinsic spread of colors across individual sources in each bin.

We adopted different stacking procedures for the \textit{HST} and for the IRAC bands, due to the different S/N available for the photometric measurements. The higher S/N available for the \textit{HST} photometry limits the measurement scatter around the true value. This allowed stacking of the \textit{HST} bands to be performed directly with the photometric measurements. The UDS and COSMOS fields lack coverage in several \textit{HST} bands compared to the two GOODS fields. Therefore, for each stack we only derived flux densities for those bands where $>90\%$ of sources had coverage in the four fields. Nonetheless, there is a set of bands common to all stacks:   $V_{606}$, $J_{125}$ and $H_{160}$. S2, S3 and S4 have measurements in the $i_{775}$ and $z_{850}$, while S3 and S4 also have $Y_{105}$. The lack of uniform coverage in the optical bands does not directly affect our analysis because at $z\sim8$ we do not expect to register any significant signal in the optical \textit{HST} filters. The added bands do add robustness to our $z\sim8$ selections in that they provide stronger discrimination against red lower-redshift sources. $JH_{140}$ covers the rest-frame UV at $z\sim8$. However, its broad wavelength transmission significantly overlaps ($\gtrsim50\%$) with both $J_{125}$ and $H_{160}$, making its use only of marginal value. The $Y$ band is valuable for constraining the redshift as the Lyman Break crosses it over the redshift range $z\sim6.4-8.7$. Nonetheless, the lack of $Y$-band coverage for all stacks is not a limitation since it is not used to determine physical parameters. The $Y$-band flux density is dependent on the exact redshift and can not be a reliable determinant of intrinsic fluxes and slopes. Regardless of filter coverage, all available bands are used to determine redshifts.

\begin{figure*}
\includegraphics[width=18cm]{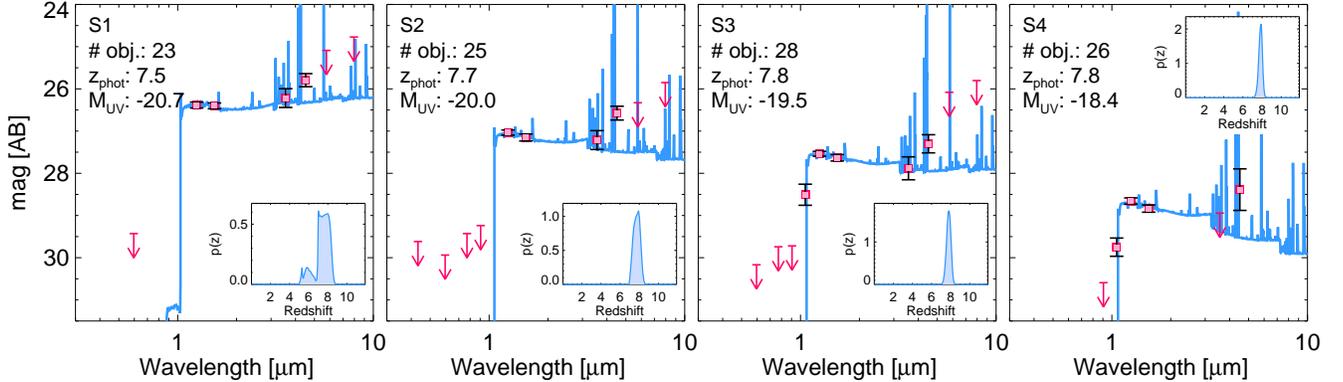}
\caption{Median-stacked SEDs resulting from our analysis. Each panel refers to a stack at a different luminosity, as indicated by the label in the top-left corner of each panel. In each panel, the filled red squares with error bars mark the stacked photometry, while the red arrows represent $2 \sigma$ upper limits. The blue curves correspond to the best-fitting \textsc{EAzY} template. The insets present the redshift probability distribution computed by \textsc{EAzY}. In the top-left corner of each panel  also indicated  are the number of objects entering the specific luminosity bin, the median redshift, and the $M_\mathrm{UV}$ computed by $\textsc{EAzY}$. All SEDs show red $[3.6]-[4.5]>0$ colors; strikingly the SEDs of the three less luminous stacks show blue $H_{160}-[3.6]<0$ colors, suggesting young stellar populations.    \label{fig:stacked_seds}}
\end{figure*}

For the IRAC stacks, the lower S/N of the IRAC bands, instead, could introduce a larger scatter into the stacked measurements if the \textit{HST} procedure was used. This potentially could affect the median estimates. Given this, stacking in the IRAC bands was performed on the actual data in cutouts centered at the nominal position of each source and cleaned of contamination from neighbours using the procedure of \citet{labbe2006, labbe2010a, labbe2010b,labbe2013, labbe2015}. The cutouts were aligned to a common location using two approaches that depended on the measured S/N of the individual object in the specific band. For $S/N \ge 5$, the location of the source was determined by  fitting a gaussian profile, re-centering the cutout to the new position and repeating this operation a second time. For $S/N<5$ instead, the cutouts were aligned to correspond to the nominal location of the source on the $H_\mathrm{160}$ mosaic, assuming no significant offset between the position on the $H_\mathrm{160}$ and that in the IRAC band. To ascertain the robustness of our results against the statistical estimator, we also repeated our stacking analysis using an inverse-variance weighted average, whose main results are presented in Appendix \ref{app:median_avg}. Reassuringly, we do not find any significant difference between the two methods.

Stacked IRAC flux densities were measured in circular apertures of $3\farcs6$ diameter and aperture-corrected using the median PSF obtained by combining the PSFs at the location of each source. Aperture correction factors were $\sim1.4, 1.4, 1.7$ and $1.8$ for the $3.6\mu$m, $4.5\mu$m, $5.8\mu$m and $8.0\mu$m bands, respectively. Flux uncertainties in the IRAC bands were computed  by bootstrapping the sample 1000 times, similarly to what was done for the \textit{HST} bands.

\begin{figure*}
\includegraphics[width=18cm]{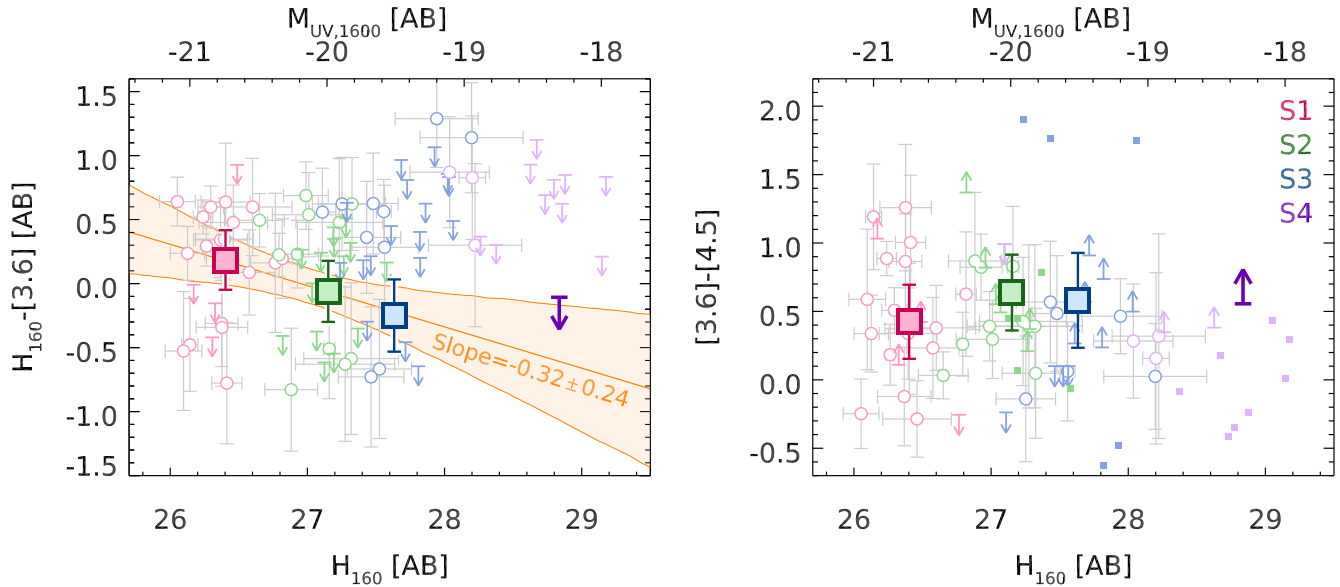}
\caption{Color-magnitude diagrams for the stacked (colored squares with error bars) and full sample (colored circles with error bars for measurements $>2\sigma$; arrows indicate $2 \sigma$ upper or lower limits; we do not show the error bars when S/N$<2$ in each band). Colors refer to specific absolute magnitude bin adopted for the stack: red, green, blue and purple for $M_\mathrm{UV}=-20.9$, $-20.2$, $-19.5$ and $-18.4$\,mag, respectively (S1, S2, S3 and S4, respectively).  The top axis marks UV absolute magnitudes obtained converting the observed $H_{160}$ flux assuming a redshift $z=8$. Stacks S3 and S4 have blue $H_{160} - [3.6]<0$\,mag at the $1-2\sigma$ level. When all stacks are considered, we find that the $H_{160} - [3.6]$ colors for lower luminosity galaxies are either systematically bluer than higher luminosity galaxies or show similar colors.   This is consistent with the idea that lower luminosity galaxies have younger stellar populations. All stacks instead have $[3.6]-[4.5]$ colors consistent with $[3.6]-[4.5]\sim0.5$\,mag, as expected because of the strong [\ion{O}{3}]+H$\beta$ line emission. All key numbers are given in Table \ref{tab:colors}.  \label{fig:irac_color}}
\end{figure*}

\section{Results}

\label{sect:results}

\subsection {Stacked SEDs at $z\sim8$}

\label{sect:sed8}

\begin{deluxetable}{cccc}
\tablecaption{Main colors of the stacked SEDs \label{tab:colors}}
\tablehead{\colhead{Label} & \colhead{$J_\mathrm{125}-H_\mathrm{160}$} & \colhead{$H_\mathrm{160}-[3.6]$} & \colhead{$[3.6]-[4.5]$}\\
& \colhead{[mag]} & \colhead{[mag]} &\colhead{[mag]} 
}
\startdata
S1 & $ -0.02\pm 0.10$ & $ 0.18\pm 0.23$ & $ 0.42\pm 0.27 $ \\
S2 & $ -0.12\pm 0.10$ & $-0.06\pm 0.24$ & $ 0.64\pm 0.28 $ \\
S3 & $ -0.09\pm 0.10$ & $-0.25\pm 0.28$ & $ 0.58\pm 0.35 $ \\
S4 & $ -0.18\pm 0.11$ & $ <-0.11$\tablenotemark{$\dagger$} & $ > 0.56$\tablenotemark{$\dagger$} \\
\hline
Average & $-0.10\pm0.05$ & $-0.03\pm0.14$ & $0.55\pm0.17$ \\
\enddata
\tablenotetext{\dagger}{$2\sigma$ limit}
\end{deluxetable}

Figure \ref{fig:irac_stacks} presents the final image stamps in the four IRAC bands for the four different bins in absolute magnitudes. To provide a better visual context for the data involved in our analysis, we also show median stacked images in the \textit{HST} bands, and remind the reader that the median flux densities in the \textit{HST} bands were estimated from the \textit{HST} photometry (see Section \ref{sect:stacking}).  The stacked photometry of the four SEDs is presented in Table \ref{tab:photometry}, and it is displayed in Figure \ref{fig:stacked_seds}, while in Table \ref{tab:colors} we list the main color estimates. Thanks to the increased depth of the GREATS data, the $3.6\mu$m and $4.5\mu$m stacks result in a clear detection in the three most luminous bins, with an $\sim4-7\sigma$ significance, and a $\gtrsim2 \sigma$ detection in the $4.5\mu$m band of S4. Remarkably, the exceptional depth of the GREATS mosaics enables a $\sim1.4\sigma$ measurement in the $3.6\mu$m band of S4. However, to provide more robust results, in this work we consider as non-detections those flux densities with $<2\sigma$ significance. The stacks in the $5.8\mu$m and $8.0\mu$m bands do not show any significant detection in any luminosity bin, as expected.

Table \ref{tab:phys_properties} summarizes the main properties of the four stacks. Absolute magnitudes $M_\mathrm{UV,1600}$ were computed with \textsc{EAzY}, fixing the redshift of each stack to the median redshift of the LBGs entering each group. As a sanity check, we also ran \textsc{EAzY} leaving the photometric redshift as a free parameter. The likelihood distributions of the photometric redshifts ($p(z)$) are reproduced as insets in Figure \ref{fig:stacked_seds}. The $p(z)$ present a single peak centered around $z\sim8$ increasing our confidence on the nature of galaxies that contribute to the stacks; the broader $p(z)$ of S1 and S2 is likely the result of the lack of $Y_{105}$ stacked data for those bins, as discussed in Section \ref{sect:stacking}.

\begin{deluxetable*}{cccr@{ $\pm$}rr@{$\pm$}lr@{ $\pm$}rcccr@{ $\pm$}rr@{ $\pm$}rc}
\tablecaption{Main physical properties of the stacked SEDs \label{tab:phys_properties}}
\tablehead{\colhead{Label} & \colhead{N. gal.\tablenotemark{a}} & \colhead{$z_\mathrm{phot}$\tablenotemark{b}} & \twocolhead{$M_\mathrm{UV}$}& \twocolhead{[\ion{O}{3}]+H$\beta$\tablenotemark{c}} & \twocolhead{H$\alpha$\tablenotemark{d}}  & \colhead{$\log(\mathrm{age}/\mathrm{yr})$} & \colhead{$A_V$} & \colhead{$\log(M_\star/M_\odot)$} & \twocolhead{SFR$_\mathrm{UV}$\tablenotemark{e}} & \twocolhead{SFR$_{\mathrm{[O\,III]}+H\beta}$\tablenotemark{f}} & \colhead{sSFR\tablenotemark{g}}\\
& & & \twocolhead{[mag]} & \twocolhead{[\AA]} & \twocolhead{[\AA]} & & [mag] &  & \twocolhead{$M_\odot$yr$^{-1}$} & \twocolhead{$M_\odot$yr$^{-1}$} & \colhead{Gyr$^{-1}$}
}
\startdata
   S1 & $ 23$ & $ 7.76$ & $ -20.7$ & $   0.1$ & $   768$ & $   279$ & $   532$ & $   193$ & $ 8.2^{+0.4}_{-0.3} $ & $ 0.3^{+0.2}_{-0.3} $ & $ 9.2^{+0.2}_{-0.2}$ & $ 23.0$ & $  1.4$ & $ 35.6$ & $ 12.9$ & $  14.9^{+10.9}_{ -6.1}$ \\
   S2 & $ 25$ & $ 7.69$ & $ -20.0$ & $   0.1$ & $  1369$ & $   389$ & $   949$ & $   270$ & $ 7.7^{+0.3}_{-1.0} $ & $ 0.1^{+0.3}_{-0.1} $ & $ 8.4^{+0.2}_{-0.5}$ & $  5.5$ & $  0.4$ & $ 17.9$ & $  5.1$ & $  21.9^{+49.1}_{ -9.2}$ \\
   S3 & $ 28$ & $ 7.75$ & $ -19.5$ & $   0.1$ & $  1184$ & $   473$ & $   821$ & $   328$ & $ 7.1^{+0.8}_{-1.1} $ & $ 0.0^{+0.2}_{-0.0} $ & $ 7.8^{+0.6}_{-0.1}$ & $  3.5$ & $  0.2$ & $  8.7$ & $  3.5$ & $  62.5^{+22.7}_{-46.9}$ \\
  S4 & $ 26$ & $ 7.70$ & $ -18.4$ & $   0.1$ &   \multicolumn{2}{l}{$>1117$\tablenotemark{$\dagger$}}     &  \multicolumn{2}{l}{$>774$\tablenotemark{$\dagger$}}  & $ 6.8^{+1.4}_{-0.8} $ & $ 0.0^{+0.2}_{-0.0} $ & $ 7.1^{+0.9}_{-0.0}$ & $  1.2$ & $  0.1$ &  \multicolumn{2}{c}{$>3.1$\tablenotemark{$\dagger$}} & $  86.7^{ +9.7}_{-76.0}$ \\
\enddata
\tablenotetext{a}{Number of galaxies in each bin of absolute magnitude.}
\tablenotetext{b}{Median photometric redshift of the sources in each bin.}
\tablenotetext{c}{Rest-frame equivalent width of [\ion{O}{3}] and H$\beta$ obtained assuming a flat $f_\nu$ SED redward of the Balmer break and the line ratios in \citet{anders2003}. These values would be $\sim30\%$ lower if we assumed that the [\ion{O}{2}] line emission only marginally contributed to the measured $3.6\mu$m-band flux density.}
\tablenotetext{d}{Rest-frame equivalent width of H$\alpha$ from the H$\beta$ estimate assuming the line ratios in \citet{anders2003} and $H\alpha/H\beta=2.85$ (\citealt{hummer1987}).}
\tablenotetext{e}{SFR computed from the rest-frame UV continuum}
\tablenotetext{f}{SFR computed from the H$\alpha$ luminosity inferred from the [\ion{O}{3}]$+H\beta$ line EW and assuming line ratios of \citet{anders2003}, with $H\alpha/H\beta=2.85$.} 
\tablenotetext{g}{sSFR based on SFR$_\mathrm{UV}$} 
\tablenotetext{\dagger}{$2\sigma$ lower limit inferred assuming  the $2\sigma$ upper limit in  the flux density of the $3.6\mu$m band.}
\end{deluxetable*}

The available filters allow, in principle, for a determination of the UV-continuum slopes $\beta$. We computed UV slopes by fitting a power law to the best-fitting SED template from \textsc{EAzY}, in a similar fashion to the procedures of \citet{finkelstein2012, duncan2014} and \citet{bhatawdekar2020}. In particular \citet{duncan2014} showed that this method is only marginally affected by bias against slopes bluer than the lower $\beta$ allowed by the templates. 
The result is that our SEDs have $\beta\lesssim-2$, broadly consistent with other recent determinations at $z\sim7-8$ (e.g., \citealt{bouwens2010, finkelstein2010, mclure2011, bouwens2014}), with a qualitative trend to bluer slopes at fainter luminosities. However, the current data make accurate determinations of the UV slopes for the stacked SEDs challenging. Indeed, the Lyman break $\lambda1215$\,\AA\ enters the $J_{125}$ band at $z\sim8.1$, decreasing the flux observed in this band, and therefore mimicking redder $\beta$ values. A simple toy model consisting of 1000 SEDs with f$_\lambda\propto \lambda^\beta$ uniformly distributed between $z=7.6$ and $z=8.6$ (i.e., for roughly half of the sample the flux in the $J_{125}$ band is affected by the Lyman break, while for the remaining half it is not) generates $\beta$ values systematically redder by $\Delta\beta\sim0.2$, for $-3.0<\beta<-1.5$. Furthermore, recent observations have shown that Lyman $\alpha$ photons may still escape the IGM even at these high redshifts (see e.g., \citealt{oesch2015, roberts-borsani2016, zitrin2015}). This can potentially increase the flux in the $J_{125}$ band leading to bluer UV slopes. Using the same toy model from above, for equivalent widths (EW$_0$) of Ly-$\alpha$ EW$_0\sim20-40$\,\AA\ (e.g., \citealt{oesch2015, roberts-borsani2016, zitrin2015, stark2017}), the UV slope becomes bluer by  $\Delta\beta\sim 0.1-0.4$. Given these systematics, we decided not to highlight the UV slopes of our stacked SEDs even though we will occasionally present some results in the context of their $J_{125}-H_{160}$ colors.

\subsection{The  $[3.6]-[4.5]$  color}
\label{sect:3645_z8}

One of the most notable features evidenced by Figures \ref{fig:stacked_seds} and \ref{fig:irac_color} is a red $[3.6]-[4.5]\sim0.5$\,mag color, which at these redshifts has been interpreted as the effect of nebular  $H\beta+$ [\ion{O}{3}] emission significantly contributing to the flux density in the $4.5\mu$m band (e.g., \citealt{schaerer2009, labbe2013}).  Figure \ref{fig:irac_color} shows that this color is approximately unchanged across $2.5$ magnitudes in UV luminosity, with the hint of a bluer color for the most luminous systems. Future higher S/N observations may well reveal changes in line ratios with luminosity.

Under the hypothesis that the measured red IRAC color is mostly the result of [\ion{O}{3}] and H$\beta$ nebular emission contributing to the flux in the $4.5\mu$m band, we can estimate  the rest-frame equivalent width (EW$_0$) of [\ion{O}{3}]$+H\beta$. For this, we assumed the line ratios of \citet{anders2003} for sub-solar metallicity ($Z=0.2Z_\odot$), and adopted a flat $f_\nu$, calibrating the $f_\nu$ continuum flux to that of the $3.6\mu$m band. At $z>7$,  [\ion{O}{2}]$_{3727}$ enters the $3.6\mu$m band. The likely strong emission by rest-frame optical lines implied by the red IRAC colors suggest that the contribution of the [\ion{O}{2}] line emission to the $3.6\mu$m-band flux density might be non-marginal. Adopting the line ratios of \citet{anders2003}, we iteratively removed the contribution of the [\ion{O}{2}] emission from the $3.6\mu$m-band flux density. We found EW([\ion{O}{2}])$\sim180-370$\,\AA, corresponding to $\sim 0.1-0.3$\,mag contribution to the $3.6\mu$m band flux density.

The resulting [\ion{O}{3}]$+H\beta$ line strengths, listed in Table \ref{tab:phys_properties}, are consistent with EW$_0($[\ion{O}{3}]$+H\beta)\sim 1000$\AA\  found at these redshifts for galaxies of comparable luminosity (\citealt{labbe2013, smit2014, smit2015, debarros2019, endsley2021}).  The largest EW$_0($[\ion{O}{3}]$+H\beta)\sim 1200-1300$\AA\  are found for S2 and S3 ($L_\mathrm{UV}\sim0.2-0.4L^*$) and are consistent with the measurements for the most extreme sources reported by \citet{endsley2021} for $L_\mathrm{UV}\sim L^*$ systems at $z\sim7$. 

Although our measurements suffer from large uncertainties, the values corresponding to S1 and S2 are consistent with the extrapolation of estimates from the $M_\star$-EW$_0$([\ion{O}{3}]$+H\beta$) found at $z\sim3.3$ by \citet{reddy2018}, in turn supporting the reduced rate of evolution with redshift  of the EW$_0$([\ion{O}{3}]$+H\beta$) observed at $z\gtrsim3$ (see e.g., \citealt{khostovan2016} and their Figure 7). 

We estimate that the [\ion{O}{3}]$+H\beta$ EW would be $\sim30\%$ smaller if the contribution of [\ion{O}{2}] to the $3.6\mu$m flux density is negligible. Although our data can not rule out this possibility, the above discussion would not change in its main conclusions, and our results would still point to exceptionally strong emission from  [\ion{O}{3}]$+H\beta$.

\subsection{The  $H_{160} - [3.6]$  color}
\label{sect:h36_z8}

At redshift $z\sim8$, the $3.6\mu$m band covers the rest-frame wavelengths just red-ward of the Balmer/4000~\AA\ break and the $H_{160}-[3.6]$ color can then provide a measurement of the strength of the break and hence an estimate of the age of the stellar populations.

\begin{figure}
\includegraphics[width=8.5cm]{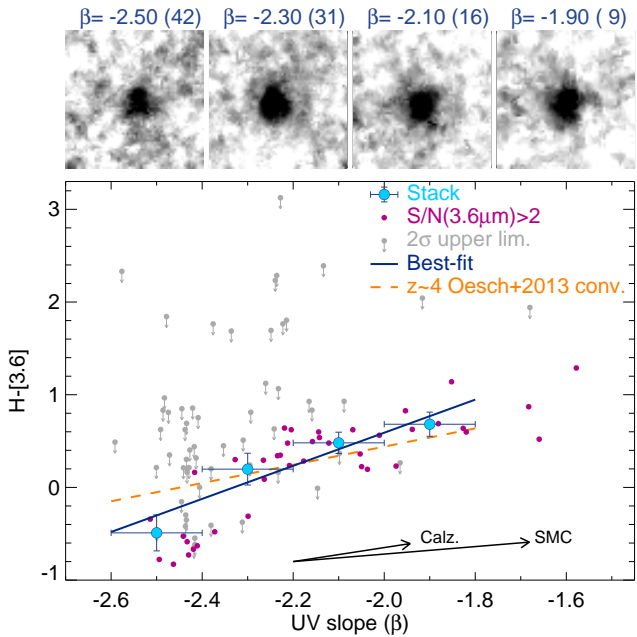}
\caption{{\bf Top Panels:} Image stamps ($\sim7\farcs2$ per side) of the $3.6\mu$m stacks in the four bins of UV slope as indicated by the label at the top of each stamp. The number of sources in each bin is shown in parenthesis. {\bf Bottom Panel:} The small filled circles and arrows correspond to the individual measurements of the $H-[3.6]$ color and $2\sigma$ upper limits, respectively, as a function of the rest-frame UV slope $\beta$ for the sample adopted for this work. The filled blue circles mark the color measured from the stacking of sources in bins of UV slope as indicated by the horizontal error bars. The best-fitting linear relation for the stacked $\beta$ values is shown by the solid blue line, while the dashed orange line represents the correlation between $J_{125}-[4.5]$ and $\beta$ found by \citet{oesch2013} at $z\sim4$ and converted to $H_{160}-[3.6]$ assuming a flat $f_\nu$ at rest-optical. The vectors at the bottom-right corner indicate the effect of a $\Delta A_V=0.1$\,mag dust extinction for an SMC law (\citealt{pei1992}) and a $\Delta A_V=0.2$\,mag for a \citet{calzetti2000} model.  \label{fig:uvslp_H36}}
\end{figure}

The $H_{160} - [3.6]$  colors are presented in the left panel of Figure \ref{fig:irac_color}, as a function of the $H_{160}$-band magnitude (where $H_{160}$ can be taken as a proxy for $M_\mathrm{UV}$ at $z\sim8$), for the individual objects in the sample and from stacking. Our stacking results show that  while $H_\mathrm{160} - [3.6] \sim +0.2 $\,mag for the most luminous stack, it becomes negative, i.e., $H_\mathrm{160} - [3.6] \lesssim 0 $\,mag  for S2, S3 and S4, indicative of very young stellar populations (more discussion about this is presented in Section \ref{sect:age}). Furthermore, our stacking results suggest a trend of redder $H-[3.6]$ with increasing luminosity, consistent with the picture of more luminous (i.e., likely more massive) systems having more evolved stellar populations. Fitting the $H_\mathrm{160} - [3.6]$ stack results vs. $H_\mathrm{160}$ magnitudes to a line, we find:
\begin{equation}
H_\mathrm{160}-[3.6]=-0.01\pm0.14-(0.32\pm0.24)(H_\mathrm{160}-27)
\end{equation}
The trend is significant at $\sim80\%$ confidence, and it is qualitatively consistent with results at lower redshifts (e.g., \citealt{gonzalez2012}), where more luminous galaxies show larger Balmer breaks than lower luminosity sources.

A number of studies have shown that the  UV slope ($\beta$) correlates with the UV luminosity of LBGs at $z>4$ (e.g., \citealt{finkelstein2012, bouwens2014,  duncan2014, bhatawdekar2020}). In Figure \ref{fig:uvslp_H36} we explore in more detail the dependence of the $H-[3.6]$ color on $\beta$, for the $\beta$ stacks and for individual sources. As noted earlier, the $\beta$-binned stacks were generated similarly to the luminosity-binned stacks (the $\beta\sim-1.7$ bin is not shown here). 

The distribution of measurements in Figure~\ref{fig:uvslp_H36} does not show any significant accumulation at the blue end, suggesting that our $\beta$ estimates have same validity. Our individual measurements indicate a prevalence of $\beta<-2$, qualitatively consistent with the blue values of the rest-frame UV found for S2, S3 and S4.

The stacked $H_{160}-[3.6]$ colors present a clear trend, becoming bluer for sources with bluer UV slopes.  A simple fit results in the following relationship:

\begin{align}
H-[3.6] &=(0.23\pm0.08)+(1.78\pm0.36)(\beta+2.2)   
\label{eq:h36}
\end{align}
valid for $-2.6\le \beta \le-1.9$. In the same figure, we also present the linear relation found by \citet{oesch2013} between $\beta$ and the $J_{125}-[4.5]$ color, from a sample of $z\sim4$ LBGs in the GOODS fields. We converted the $J_{125}-[4.5]$ color into a $H_{160}-[3.6]$ color assuming a flat $f_\nu$ continuum red-ward of the Balmer break, and a power law with slope $\beta$ for the rest-frame UV. Our linear relation (Eq. \ref{eq:h36}) is a little steeper than that of \citet{oesch2013}, indicative of potential evolution towards redder $H_{160}-[3.6]$ colors between $z\sim8$ and $z\sim4$, qualitatively consistent with an ageing of the  stellar populations with cosmic time. However, a closer inspection suggests that the steeper slope we observe at $z\sim8$ is also likely driven by the very blue $H_{160}-[3.6]\sim-0.5$\,mag for $\beta\sim-2.5$, as the measurements for $\beta\ge-2.3$ are consistent at $\gtrsim1\sigma$ with the $z\sim4$ relation.

Accounting for a $\sim0.2$\,mag contribution from [\ion{O}{2}] line emission to the $3.6\mu$m band, as we derived in Section \ref{sect:3645_z8}, would bring our stacked measurements for the $H-[3.6]$ color as a function of $\beta$ into even better agreement with those of \citet{oesch2013} for $\beta>-2.3$, if we assume that the $z\sim4$ relationship is only marginally affected by nebular line emission. Nonetheless, at $z\sim4$, \citet{faisst2016} found EW($H\alpha$)$\sim300$\AA, resulting in $\lesssim0.1$\,mag contribution from [\ion{O}{2}], supporting our initial assumptions.

\section{Discussion}
\label{sect:discussion}

\subsection{Age and $M_\star/L_\mathrm{UV}$ ratio of the stacked SEDs}
\label{sect:age}

\begin{figure*}
\includegraphics[width=18cm]{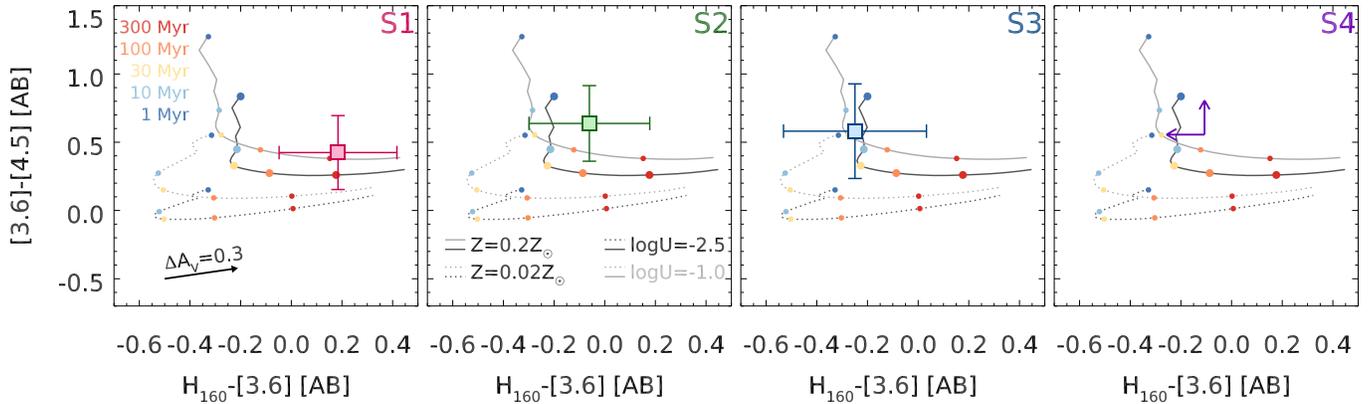}
\caption{Comparison of the stacked results to models using the $[3.6]-[4.5]$ vs $H_{160}-[3.6]$ plane. Each panel refers to one of the stacks, as indicated by the labels at the top-right corners. In each panel we also present tracks from the synthetic SEDs with a constant star-formation history: solid curves mark tracks for a $Z=0.2Z_\odot$ metallicity, while the dotted curves are for $Z=0.02Z_\odot$. We also distinguish between two different ionization parameters, with the black curves marking $\log U=-2.5$, while the grey curves correspond to $\log U=-1$. The colored filled circles mark ages on approximately a logarithmic scale, as indicated by the legend at the top-left corner in the left-most panel, while  the black arrow shows the effect of increasing the extinction by $0.3$\,mag. Other plotting conventions are as in Figure \ref{fig:irac_color}. Ages younger than $\sim 100$\,Myr are preferred for all but the most luminous stack. Older ages from lower metallicities are ruled out by the implied bluer $[3.6]-[4.5]$ colors resulting from lower EW [\ion{O}{3}]$_{4959,5007}$ emission.\label{fig:model_colors}}
\end{figure*}

In Figure \ref{fig:model_colors} we compare the observed color $[3.6]-[4.5]$ vs. $H_{160}-[3.6]$ to synthetic colors extracted from a library of SED templates based on the \citet{bruzual2003} set. We considered templates with constant  SFH and metallicity $Z=0.2 Z_\odot$. They were processed though \textsc{cloudy} version 17.02 \citep{ferland2017}, assuming a spherical constant-density nebula with $n(H)=100$\,cm$^{-3}$, same gas metallicity as the stellar component, an ionization parameter $\log U=-2.5$, consistent with recent results at similar redshifts (e.g., \citealt{stark2017,debarros2019}), and assuming the escape fraction to be negligible.

The blue $H_{160}-[3.6]\lesssim 0$\,mag of S2, S3 and S4 indicate ages younger than $\sim100$\,Myr, while the redder color for S1 suggests ages $\sim100-300$\,Myr. These young ages are also supported by the extreme red $[3.6]-[4.5]\sim0.5$\, mag colors, indicative of strong line intensities.

It is well established (e.g., \citealt{renzini1986}) that the broad-band SED of a young, metal-rich system can mimic an older, metal-poor one, and vice versa. Specifically, in our case, metallicities lower than $Z=0.2 Z_\odot$ could imply stellar ages older than $\sim100$ Myr. To explore this scenario, in Figure \ref{fig:model_colors} we also present synthetic tracks for $Z=0.02 Z_\odot$ and $\log U=-2.5$. The lower metal abundances result in decreased intensity of the optical emission lines. The corresponding bluer $[3.6]-[4.5]$ colors are inconsistent with our measurements. Addition of $\gtrsim0.4$ mag dust extinction would bring the track closer to our measurements, but with ages $\lesssim 100$ Myr. Finally, because the ionization parameter can affect the intensity of emission lines, we also consider the effects of an exceptionally intense $\log U=-1.0$ ionizing field.  The corresponding track is consistent with our estimates for very young ages ($<$few$\times$ Myr), while the addition of dust extinction would result in ages younger than $\sim 100$ Myr.

The observed trend towards redder $H_{160} - [3.6]$ colors at higher luminosities suggests that we could be witnessing the development of a more evolved stellar population at just $\sim 650$\,Myr of cosmic time, indicative that the ages of the stellar populations could depend on mass already at such early epochs. The degree of such evolution is hard to establish though, since bright young stars could be outshining any underlying older stellar populations (e.g., \citealt{papovich2001}). Recent work suggests the existence of evolved stellar populations in $L\lesssim 1-0.4 L^*$ individual galaxies identified at  $z\sim8-10$  (e.g., \citealt{salmon2018, hoag2018, hashimoto2018, roberts-borsani2020, strait2020}). Indeed, our data also show $H_{160}-[3.6]>0$\,mag for some of the sources, supporting the existence of a more evolved stellar population even for $L<L^*$. Our stacking results, however, indicate that these more evolved populations do not constitute the bulk of the galaxy population at these redshifts, likely implying large duty cycles of star formation. Better quantifying the duty cycle would require the direct detection of a higher fraction of sources in the $3.6\mu$m band. For this we await JWST.

In Table \ref{tab:phys_properties} we present the main stellar population parameters for the stacks we obtained  running \textsc{FAST} with the \citet{bruzual2003} template set and adding nebular continuum and emission line contribution with \textsc{Cloudy} and the same configuration described above. To reduce a potential bias towards extremely young ages, the fitting for S1, S2 and S3 was performed on the photometry cleaned of the emission lines contribution, using the information on the EW we obtained in Section \ref{sect:3645_z8}. For consistency, emission lines were also removed from the SED templates. Physical parameters for S4 were determined by adopting the pristine photometry with templates including emission lines, given the $1.4\sigma$ detection in the $3.6\mu$m band. The results show that stacks have mean stellar masses per source in the range $M_\star\sim10^{7.1-9.2}M_\odot$, implying mass-to-light ratios $M_\star/L_\mathrm{UV}\sim 0.003-0.04 M_\odot/L_\odot$, consistent with values found at similar redshifts for $L\sim1.4-3.6L^*$ sources (e.g., \citealt{stefanon2019}).

\subsection{Star-Formation Rate}

Following the relation between rest-frame UV luminosity and SFR  of \citet{madau2014}  for sub-solar metallicity ($\log(Z/Z_\odot)=-0.5$) and no dust attenuation,  we calculate for our sample  SFR$_\mathrm{UV}\sim1-23 M_\odot$/yr (see Table \ref{tab:phys_properties}).

We can also estimate the SFR from the nebular emission, using the results on the EW$_0$ of [\ion{O}{3}]$+H\beta$ we derived in Section \ref{sect:3645_z8}, and which we denote with SFR$_{\mathrm{[O\,III]}+H\beta}$. Assuming Case B recombination, the line ratios of \citet{anders2003} for $Z=0.2Z_\odot$ and H$\alpha/$H$\beta=2.85$ (\citealt{hummer1987}), we  converted the measured equivalent width of [\ion{O}{3}]$+H\beta$ into $H\alpha$ line intensity. The \citet{kennicutt2012} relation for the $H\alpha$ luminosity then gives SFR$_{\mathrm{[O\,III]}+H\beta}\sim 3-36 M_\odot$/yr. The individual measurements are listed in Table \ref{tab:phys_properties}.

\begin{figure}
\includegraphics[width=8.6cm]{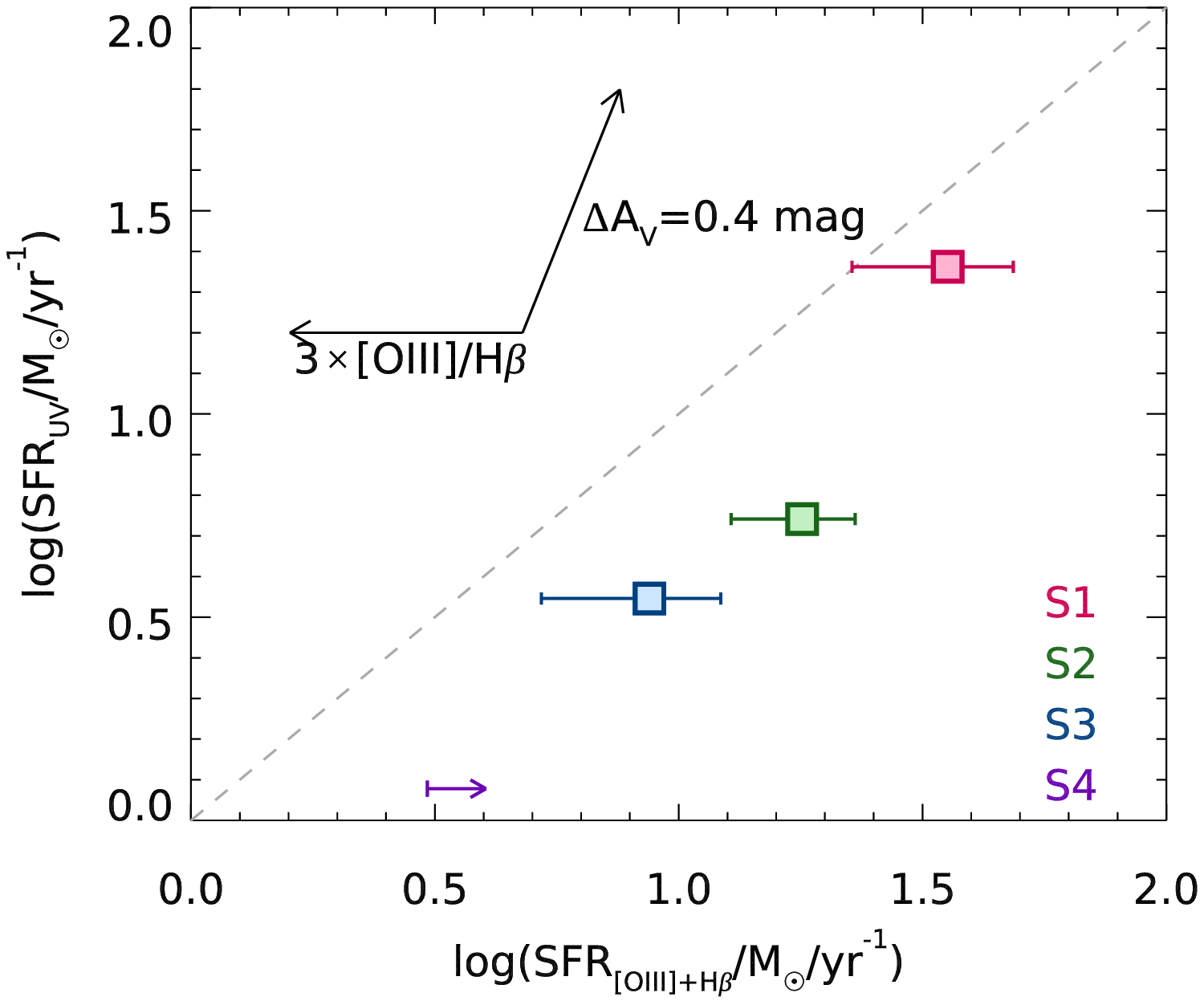}
\caption{Comparison between SFR$_\mathrm{UV}$ and SFR$_{\mathrm{[O\,III]}+H\beta}$. The former is computed on the basis of the UV luminosities of galaxies, and the latter is computed converting the [\ion{O}{3}]+H$\beta$ luminosity measured from the IRAC photometry into $H\alpha$ luminosity assuming the line ratios of \citet{anders2003} and $H\alpha/H\beta=2.85$. The $H\alpha$ luminosity is finally converted into SFR using the relation of \citet{kennicutt2012}. The colored points with error bars or lower limits refer to the four stacks as indicated by the legend. The grey dashed line marks the identity,  SFR$_{\mathrm{[O\,III]}+H\beta}=$ SFR$_\mathrm{UV}$. The vectors at the top-left corner show the impact of increasing the [\ion{O}{3}]/H$\beta$ ratio by $3\times$, and correcting for a \citet{calzetti2000} dust extinction of $\Delta A_V=0.4$\,mag.  \label{fig:sfr_uv_ha}}
\end{figure}

In Figure \ref{fig:sfr_uv_ha} we compare the SFR$_{\mathrm{[O\,III]}+H\beta}$ to SFR$_\mathrm{UV}$. While for S1 the two measurements are consistent at $\gtrsim1\sigma$, the SFR$_{\mathrm{[O\,III]}+H\beta}$  for S2 and S3 are  $\sim3-5\times$ higher than the SFR$_\mathrm{UV}$. A similar result (SFR$_{H\alpha}$/SFR$_\mathrm{UV}\sim6$) was found by \cite{shim2011} from analyzing the observed blue IRAC $[3.6]-[4.5]$ color for a spectroscopically-confirmed sample of LBGs at  $3.8<z<5.0$ over the GOODS fields  (see also \citealt{smit2015, shivaei2015}). These studies analyzed different mechanisms to explain such discrepancy, such as dust attenuation,  top-heavy IMFs or increasing ionization parameter, even though no dominant process has been identified so far. In the following paragraphs we discuss three potential factors that could explain or at least in part contribute to the systematic difference in the SFR estimates from the UV continuum and $H\alpha$ emission: star-formation history, dust extinction and line ratios evolving with time. However, the relative contribution of each of these three scenarios is difficult to ascertain, and roughly comparable contributions may also constitute a plausible scenario.

Based on our results in the previous section, the stacked SEDs are likely characterized by very young ages ($\lesssim 30-100$\,Myr). The UV stellar continuum is mostly generated by O and B stars with typical lifetimes of $\sim 100-200$\,Myr (\citealt{kennicutt2012}).  A constant SFR equal to the estimated SFR$_\mathrm{UV}$ imply that those stellar masses could be generated in just $\sim15-50$\,Myr, i.e. we might still be witnessing the first burst of star-formation, started at $z\sim 9$. However, because the $H\alpha$ emission will be the result of more recent star formation ($\lesssim 10$ Myr - e.g., \citealt{kennicutt2012}),  the higher SFR$_{\mathrm{[O\,III]}+H\beta}$ values we inferred may indicate a SFR increasing with time. This arises because the stacking analysis, by construction, lessens the variation between individual measurements.

Systematic differences in SFR measurements could be introduced by the differential nature of dust extinction with wavelength. Assuming a \citet{calzetti2000} extinction curve and assuming that the extinction curve of the nebular emission is the same of the stellar continuum ($E(B-V)_\mathrm{neb}=E(B-V)_\mathrm{cont}$ - e.g., \citealt{erb2006b, reddy2012} - but see e.g., \citealt{calzetti2000, steidel2014, reddy2020} for different results), we find that an extinction of $A_V\sim0.4-0.5$\,mag is required in order to match the two SFR estimates. This extinction would result in UV slopes redder by $\Delta\beta\sim0.7$, or $\beta\sim-2$ (when assuming the bluest intrinsic slope value to be $\beta=-2.6$, as we measure for the youngest template), qualitatively only marginally consistent with the upper limit of the UV slopes of our stacks.

A third possibility for explaining the systematic offset between the SFR$_{\mathrm{[O\,III]}+H\beta}$ and SFR$_\mathrm{UV}$ is if the nebular line ratios are evolving with cosmic time. Indeed, an increase in the ionization parameter (e.g., harder spectrum) has been suggested by recent work (e.g., \citealt{maseda2020}). This can result in an increase of the   [\ion{O}{3}]/H$\beta$ ratio  (see e.g., discussions in \citealt{brinchmann2008} and \citealt{steidel2014}).

The [\ion{O}{3}]$\lambda5007$/H$\beta=4.7$ adopted in our estimates (\citealt{anders2003}) is consistent with recent determinations from spectroscopic measurements of LBGs at $z\sim2-4$ (e.g., \citealt{holden2016, dickey2016}).  Forcing the relation SFR$_{\mathrm{[O\,III]}+H\beta} =$ SFR$_\mathrm{UV}$, and keeping [\ion{O}{3}]$\lambda5007$/[\ion{O}{2}] $=3$, the red $[3.6]-[4.5]$ color implies [\ion{O}{3}]$\lambda5007/H\beta = 17\pm 8$ and $13\pm8$, for S2 and S3, respectively. This is $\sim3\times$  the ratio we initially adopted (see Figure \ref{fig:sfr_uv_ha}). 

Even though estimates of the [\ion{O}{3}]$\lambda5007/H\beta$ ratio are still very uncertain at high redshifts, a number of studies suggest the existence of a trend of increasing [\ion{O}{3}]/H$\beta$ with redshift (e.g., \citealt{brinchmann2008, steidel2014, kewley2015, faisst2016, harikane2018b}). In particular, \citet{faisst2016} found that at $z\gtrsim5$ the [\ion{O}{3}]/H$\beta$ ratio could be larger,  [\ion{O}{3}]/H$\beta\gtrsim3$, with values potentially as high as [\ion{O}{3}]/H$\beta\sim15$ and with a lower limit of  [\ion{O}{3}]/H$\beta\sim2$. On the other hand, using a spectroscopically confirmed sample of Ly$\alpha$ emitters at $z\sim5.7, 6.6$ and $7.0$, \citet{harikane2018b} found that the ratio [\ion{O}{3}]/H$\beta$ depends on the Ly$\alpha$ EW but it is roughly independent from redshift (see their Figure 15). Specifically, at $z\sim5.7$, which provided the widest coverage in terms of EW$_{\mathrm{Ly}\alpha}$, they measured a maximum value of [\ion{O}{3}]/H$\beta\sim10$, for EW$_{\mathrm{Ly}\alpha}\sim90$\AA, consistent with our estimates for S2 and S3. However, their measurement at $z\sim7$ results in an upperlimit ([\ion{O}{3}]/H$\beta<2.8$) somewhat in tension with our results.

\subsection{Specific star-formation rate and halo mass growth}

\begin{figure*}
\includegraphics[width=18cm]{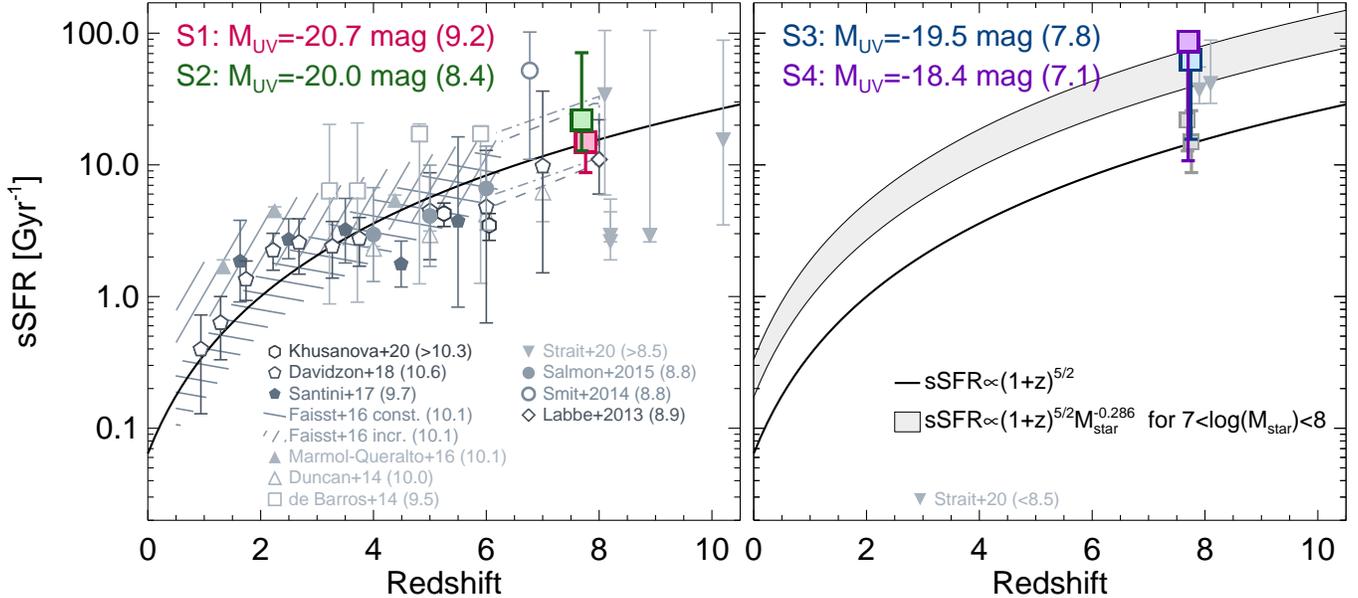}
\caption{Evolution of the sSFR with redshift. The measurements from the stacking analysis are presented in the panels according to their $M_\mathrm{UV}$, indicated at the top (together with our "S" label). For reference, in the right panel we also reproduce our measurements for the two more luminous bins (grey squares). In each panel we also plot recent estimates of the sSFR  at high redshifts from the literature, as indicated by the legends at the bottom, segregated according to their stellar mass, quoted in parenthesis in log scale. The solid black curve corresponds to the evolution of the sSFR from the toy model of \citet{dekel2013}, of the form sSFR $\propto (1+z)^{5/2}$, as expected from cold gas inflow that follows the hierarchical merging of the dark matter halos. The filled grey region marks the \citet{dekel2013} model, renormalized to $M_\star=10^{7}-10^{8}M_\odot$ using the \citet{reddy2018} stellar mass scaling for  $H\alpha$ and assuming the \citet{dekel2013} relationship to be representative of $M_\star=10^{9.5}M_\odot$ galaxies. The overall good match of the curves to the observations suggests there is no strong evolution in the star-formation efficiency of galaxies with cosmic time. \label{fig:sSFR}}
\end{figure*}

Combining our results from the previous sections we can now estimate the sSFR at $z\sim8$. The resulting values are listed in Table \ref{tab:phys_properties} and shown in Figure \ref{fig:sSFR}, segregated according to the UV luminosity (which can roughly be taken as a stellar mass proxy). These values are consistent with the young ages we previously recovered, as for a constant SFH, the sSFR$\sim 1/$age$\sim 6-160$\, Gyr$^{-1}$.  It is perhaps noteworthy that the sSFR we derive at different stellar masses are consistent with each other. While recent observations indicate that the sSFR marginally depends on stellar mass to $z\sim6$ for $\log(M_\star/M_\odot)\lesssim 10$ (e.g., \citealt{fumagalli2012,whitaker2014, schreiber2015, salmon2015, reddy2018},  current models still provide contrasting predictions for a mass-dependent sSFR (see e.g., \citealt{behroozi2013}) or for a mass-independent sSFR (see e.g., \citealt{sparre2015, ceverino2018}). The agreement between measurements at different stellar masses could be interpreted as  the existence of a main sequence of star-forming galaxies across $1.5$ orders of magnitude already at $\sim650$\, Myr of cosmic time.

In Figure \ref{fig:sSFR}  we also compare our sSFR estimates to recent determinations at high redshifts from \citet{labbe2013, smit2014, duncan2014, salmon2015,marmol2016, santini2017, davidzon2018, khusanova2020} and \citet{strait2020}, together with the measurements at $0\lesssim z \lesssim 6$ of  \citet{faisst2016}, segregated by stellar mass and/or $M_\mathrm{UV}$. Most of theses measurements correspond to stellar masses $>10^9 M_\odot$ and show large systematic differences for $z>4$, spreading over $\sim1$ order of magnitude, with sSFR ranging from $\sim3$\,Gyr$^{-1}$ (see also \citealt{endsley2021}) to $\sim 20$\,Gyr$^{-1}$.

Our estimates are consistent with those of $z\sim7-8$ LBGs from broad band photometry (\citealt{ labbe2013, smit2014, jiang2016, castellano2017}) and with the extrapolation \citet{faisst2016} propose to $z\sim8$. They indicate an increasing sSFR with redshift out to $z\sim8$, irrespective of the stellar mass bin.\newline

The solid black curve in Figure \ref{fig:sSFR} marks the evolution of the sSFR from the toy model of \citet{dekel2013}. In this model, the specific dark matter halo accretion rate (SHMAR) evolves as $\propto (1+z)^{5/2}$. It is interesting to note that the exponent $5/2$ was obtained analytically from simple theoretical considerations on the Extended Press-Schechter formalism describing the evolution of the dark matter halo assembly (see also \citealt{neistein2008b, weinmann2011, genel2014}). The SHMAR was then converted into sSFR assuming  that the baryonic accretion involved exclusively cold gas. The resulting form is simply a re-normalization of the SHMAR, which maintains the dependence of $\propto (1+z)^{5/2}$. Similarly, a sSFR increasing with redshifts out to $z\sim8$ was also predicted by \citet{tacchella2013} assuming the SFH to closely follows the halo assembly time,  by \citet{ceverino2018}, with which our measurements are consistent to within a factor of 2 and by \citet{park2019} which assumed a non-evolving star-formation efficiency.

The similarity between the evolution of the sSFR and of the SHMAR  at  $z\gtrsim3 $ suggests that at high redshifts  galaxy formation could be dominated by the assembly of cold gas, driven by the hierarchical formation of the dark matter halos, with marginal dependence on an evolving star-formation efficiency. This result is qualitatively consistent with a marginal evolution with redshift of the $M_\star/M_h$ ratio recently observed at $z>4$ (\citealt{stefanon2017b, stefanon2021b})  and with such models providing a natural explanation for the evolution in the UV LF at $z\gtrsim4$ (\citealt{tacchella2013, tacchella2018, bouwens2015, bouwens2021, mason2015, harikane2018, park2019}).

Because feedback mechanisms could influence the star formation differently depending on the stellar and halo mass, in the right panel of Figure \ref{fig:sSFR} we also present (grey filled region) a speculative curve for the sSFR with a dependence on stellar mass. Specifically, we assumed that the \citet{dekel2013} relationship well represents galaxies with a stellar mass of $M_\star=10^{9.5}M_\odot$; this was then rescaled  for $M_\star=10^{7}-10^{8}M_\odot$ with the $M_\star^{-0.286}$ dependence of $H\alpha$ luminosity of  \citet[but see e.g., \citet{ceverino2018} who find a marginal dependence of the sSFR with stellar mass]{reddy2018}.

\section{Summary and Conclusions}
\label{sect:conclusions}

Through the analysis of the deepest available \textit{Spitzer} IRAC and \textit{Hubble} data for a significant sample of star-forming galaxies at $z\sim8$, we have been able to gain  striking new insights into the star formation history of galaxies just $650$\, Myr after the Big Bang.  These insights are derived on the basis of deep stacked SEDs of $z\sim8$ Lyman-break galaxies identified over the CANDELS/GOODS-N/S, ERS, XDF, CANDELS/UDS and CANDELS/COSMOS fields.  These fields are characterized by deep coverage with \textit{HST}  in the \textit{HST}/ACS $V_\mathrm{606}$ and  $I_\mathrm{814}$, \textit{HST}/WFC3 $Y_\mathrm{105}$, $J_\mathrm{125}$, $JH_\mathrm{140}$ and $H_\mathrm{160}$ bands, and by \textit{Spitzer}/IRAC in the $3.6\mu$m and $4.5\mu$m bands. In particular, the GOODS fields benefit from ultradeep IRAC $3.6\mu$m and $4.5\mu$m mosaics, combining all the observations from the recently completed GOODS Re-ionization Era wide-Area Treasury from Spitzer - GREATS; \citealt{stefanon2021a}). This program brings near-homogeneous $\sim200-250$~hr depth (corresponding to $5 \sigma \sim 26.8-27.1$~mag) in \textit{Spitzer}/IRAC $3.6\mu$m and $4.5\mu$m in 200 arcmin$^2$ over the two GOODS fields.

The full sample was segregated into four UV luminosity bins (S1, S2, S3 and S4, in order of decreasing luminosity). Stacking in the \textit{HST} bands was performed directly on the flux measurements, while for the IRAC bands we carried out aperture photometry on the stacked image cutouts, after each source was cleaned of neighbouring sources. We excluded those cutouts with neighbouring residuals that remained from the \textsc{Mophongo} image processing that overlapped with the nominal location of a source. To account for the overall shape of each SED, stacking was performed after normalizing the fluxes and cutouts of each source to the weighted mean of the measurements in the $J_{125}$ and $H_{160}$ bands.

Our main observational results can be summarized as:

\begin{itemize}
\item The stacked SEDs are characterized by $H_{160} - [3.6]$ colors ranging between $\sim-0.3$\,mag and $\sim+0.2$\,mag. The blue $H_{160} - [3.6]\lesssim0$\,mag colors we measure for the three faintest stacks imply very young stellar ages ($\lesssim10^8$\,yr).
\item We explored the dependence of the $H_{160}-[3.6]$ color on the rest-frame UV luminosity. We find that our measurements are consistent with either no trend or for fainter galaxies to have younger ages (Figure \ref{fig:irac_color}). This suggests that the stellar population ages of galaxies may be mass-dependent $\sim650$\,Myr after the Big Bang.
\item The stacks are characterized by red $[3.6]-[4.5]\gtrsim0.5$\,mag colors. Given the plausible assumption that the red $[3.6]-[4.5]$ color results from the [\ion{O}{3}] and $H\beta$ emission lines entering the $4.5\mu$m band at $z\gtrsim7.5$, these colors correspond to EW$_0$([\ion{O}{3}]+$H\beta$)$\sim800-1300$\AA\ (Table \ref{tab:phys_properties}). 
\end{itemize}

The above results lead to the following conclusions:

\begin{itemize}
\item The SFR we infer from the measured [\ion{O}{3}]$+H\beta$ EWs are factors $\sim3$ higher than the SFR derived  from the UV luminosity.   Possible explanations include either an increasing SFR with time or an [\ion{O}{3}$\lambda5007$]/$H\beta$ ratio evolving with time as a result of an evolving ionizing radiation field (Figure \ref{fig:sfr_uv_ha}). A dust extinction of $A_V\sim0.4$ mag is less effective as an explanation, given that it would imply UV slopes $\beta$ redder by $\Delta\beta\sim0.7$, only marginally consistent with our current and other earlier observations.  An older stellar population also is not effective as an explanation, given the clear discrepancy with the measured  blue $H_{160} - [3.6] \lesssim 0$ mag colors.
\item The high SFRs inferred from the UV light imply sSFRs of $\gtrsim10$\,Gyr$^{-1}$. When these new  sSFR values at $z\sim8$ are compared to the sSFR at lower redshifts, they suggest a very similar evolution in the sSFR of galaxies and the specific halo mass accretion rate, and, ultimately, a marginally evolving $M_\star/M_\mathrm{h}$ ratio for $z\gtrsim3$. However, the sSFR for the two fainter luminosity bins are also consistent with a scenario where the sSFR weakly depends on stellar mass (Figure \ref{fig:sSFR}).
\end{itemize}

The results presented in this paper are based on broad band  photometry probing  rest-frame wavelengths up to $\sim0.6\mu$m. More accurate estimates of the stellar continuum red-wards of the Balmer Break will benefit by probing the continuum SED of galaxies at $>6\mu$m. Upcoming facilities like JWST will be pivotal in placing these results on a more robust observational footing. Specifically,  the increased sensitivity at redder wavelengths  will allow us to probe the contribution of lower-mass evolved stars which dominate the stellar mass. Observations with higher spectral resolution will, on the other hand, allow us to further assess the ionization state of these systems and their star-formation histories.

\appendix

\section{Comparison to previous SEDs at $z\sim8$}
\label{sect:sed_comparison}

Given the challengingly low S/N  in the IRAC bands for $z\sim8$ LBGs available so far, only a limited number of studies have been undertaken that investigate  the average stellar population properties of $z\sim8$ LBGs. 

\citet{labbe2013} leveraged deep IRAC $3.6\mu$m and $4.5\mu$m imaging from the IUDF program (\citealt{labbe2015}) to explore $z\sim8$ LBGs with a sample of 76 $Y-$dropouts. Their sample has luminosities that correspond to our S1 and, in part, S2. The \citet{labbe2013} stacked colors for $H_{160}-[3.6]\sim0.4-0.5$\,mag suggest more evolved stellar populations than our results indicate. Reassuringly, the median of our $H_{160}-[3.6]$ measurements in our work for sources in common between the two studies is $H_{160}-[3.6]\sim0.3$\,mag. Our HST photometry was calibrated to be statistically consistent with the 3D-HST values. Comparison to  matching sources in the 3D-HST catalog (\citealt{skelton2014, momcheva2016}) also showed that the $H_{160}$-band fluxes of \citet{labbe2013} are fainter by approximately $0.15$\,mag. Complicating the comparison, we found that the sources in common between the two samples in our analysis are split into two distinct magnitude bins,  and that our sample also includes sources with bluer $H_{160}-[3.6]$ colors than found by \citet{labbe2013} for the same range of magnitudes.

\citet{song2016} discussed $z\sim8$ median SEDs from stacking the photometry of 77 sources at $7.5<z_\mathrm{phot}<8.5$ from the \citet{finkelstein2015a} sample over the CANDELS GOODS fields.  Their brightest bin ($M_\mathrm{UV}\sim-20.95$\,mag) exhibits a moderately red color $H_{160}-[3.6]\sim 0.3$\,mag, consistent with our estimate for S1. However, the lack of detection in the $4.5\mu$m band, and in the other IRAC bands for lower luminosity bins, is likely the result of the shallower depth of the IRAC data available at that time (before GREATS), making further comparisons more difficult.

\section{Comparison between median and inverse-variance weighted average stacked SEDs}
\label{app:median_avg}

\begin{figure}
\includegraphics[width=9.3cm]{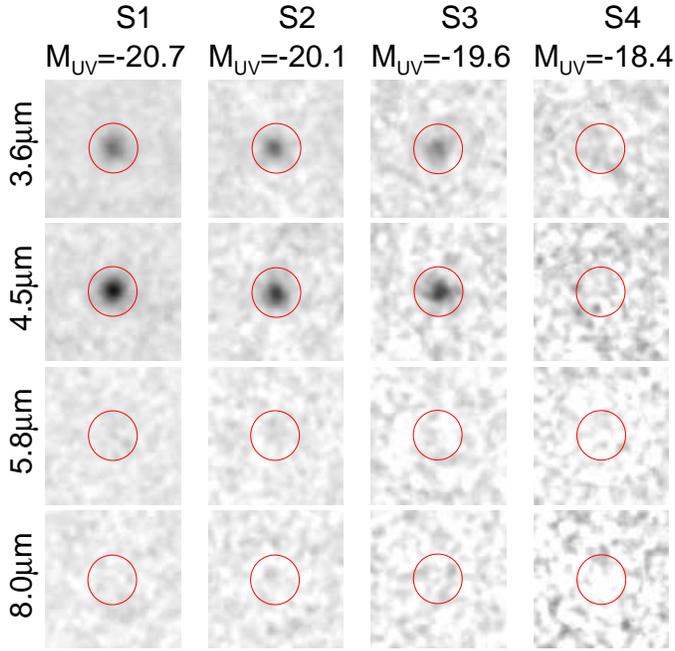}
\caption{Stacked IRAC cutouts using the inverse-variance weighted average estimator in the four bands and for the four absolute magnitude segregations, as indicated at the top of the figure. Other plotting conventions are as in Figure \ref{fig:irac_stacks}.\label{fig:app_irac_stacks}}
\end{figure}

\begin{figure*}
\includegraphics[width=18cm]{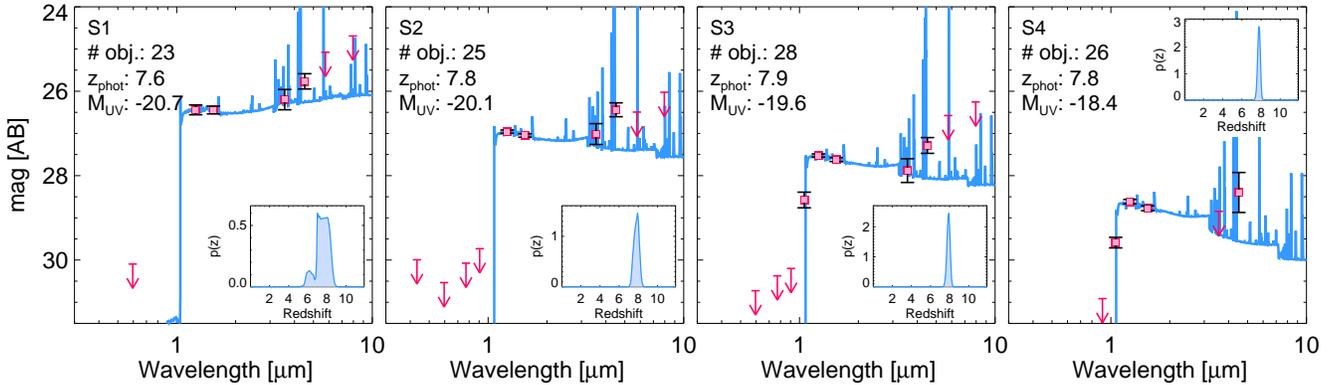}
\caption{Full stacked SEDs in the four absolute magnitude bins for the inverse-variance weighted average stacking procedure. Other plotting conventions are as in Figure \ref{fig:stacked_seds}. \label{fig:app_seds}}
\end{figure*}

We tested the robustness of the stacked SEDs for the IRAC stacks against the adopted statistical estimator by repeating our analysis with the inverse-variance weighted average instead of the median, adopting the value of the exposure time in the relevant IRAC band for the weight.

\begin{figure*}
\includegraphics[width=18cm]{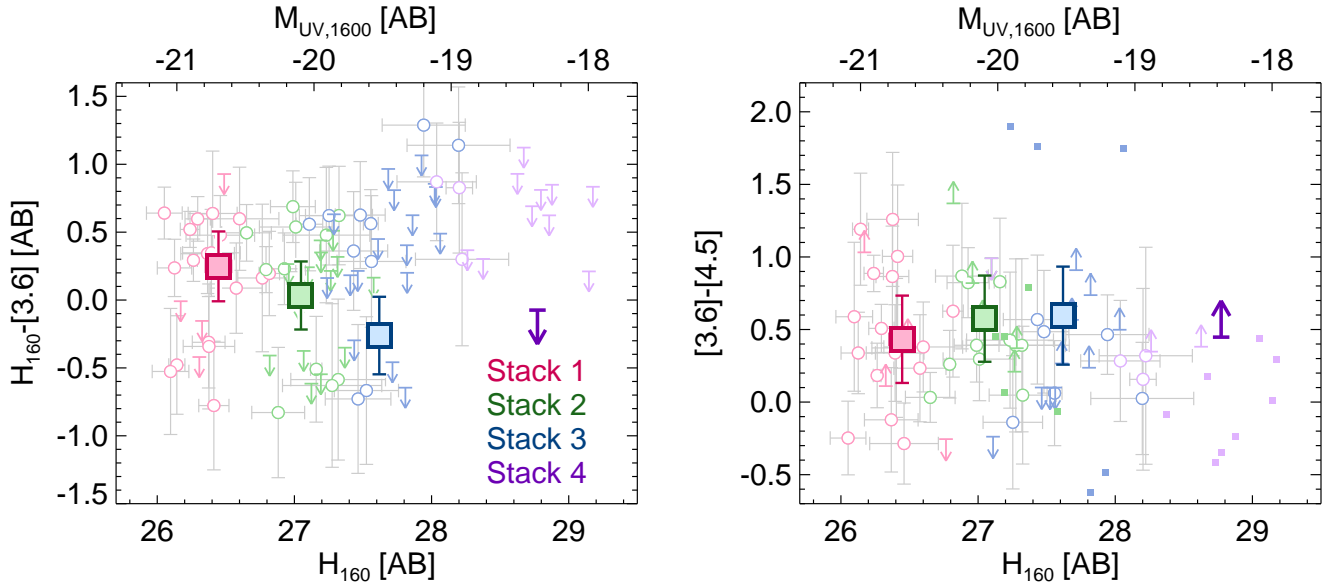}
\caption{Color-magnitude plots for the inverse-variance weighted average stacks. The two panels shown here adopt the same plotting conventions as in Figure \ref{fig:irac_color}. \label{fig:app_colors}}
\end{figure*}

\begin{figure*}
\includegraphics[width=18cm]{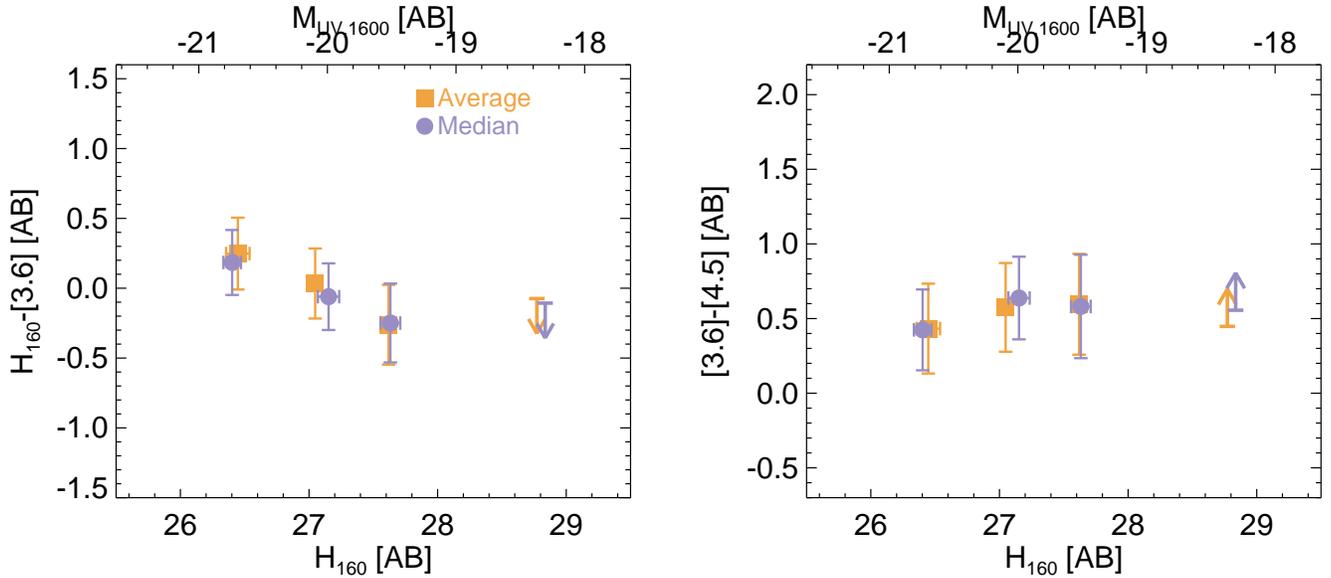}
\caption{Comparison of the colors from the inverse-variance weighted average (filled yellow squares) to those from the median (filled purple circles). The median is our primary approach in the paper. No systematic difference is observed  for any of the four bins.  \label{fig:app_avg_med_comp}}
\end{figure*}

The weighted mean stacked cutouts in the four IRAC bands are shown in Figure \ref{fig:app_irac_stacks}, while the resulting SEDs are presented in Figure \ref{fig:app_seds}. The corresponding $H_{160}-[3.6]$ and $[3.6]-[4.5]$ colors are represented in Figure \ref{fig:app_colors}, while in Figure \ref{fig:app_avg_med_comp} we compare the main colors from the inverse-variance weighted average to those from the median analysis. This last Figure shows that there is no significant difference between the two analysis. The limit shown in the $[3.6]-[4.5]$ vs. $H_{160}$ plot for S4 in Figures \ref{fig:app_colors} and \ref{fig:app_avg_med_comp} results from a $<2\sigma$ detection in the $4.5\mu$m band when the inverse variance mean is adopted.

\bibliographystyle{apj}

\end{document}